\documentclass[12pt]{elsarticle}
\usepackage{graphicx} 
\usepackage[T1]{fontenc}
\usepackage[usenames]{color} 
\usepackage{ulem}
\usepackage{amsmath,amssymb,theorem} 
\usepackage{bm}
\usepackage{dcolumn}
\usepackage{hhline}
\usepackage{float}
\usepackage{subfig}
\usepackage{lineno}
\biboptions{sort&compress}

\journal{Applied Surface Science}

\begin{document}

\begin{frontmatter}

\title{Roughness and Correlations in the Transition from Island to Film Growth:
Simulations and Application to CdTe Deposition}

\author{Tung B. T. To}
\ead{tobathanhtung@gmail.com}
\address{Instituto de F\'\i sica, Universidade Federal Fluminense,\\
Avenida Litor\^anea s/n, Niter\'oi, RJ, 24210-340, Brazil}
\author{Renan A. L. Almeida}
\ead{renan@almeidaphys.com}
\address{Instituto de F\'\i sica, Universidade Federal Fluminense,\\
Avenida Litor\^anea s/n, Niter\'oi, RJ, 24210-340, Brazil}
\address{Department of Physics, Federal University of Vi\c{c}osa, Vi\c{c}osa 36570-900, MG, Brazil}
\author{Sukarno O. Ferreira}
\ead{sukarno@ufv.br}
\address{Department of Physics, Federal University of Vi\c{c}osa, Vi\c{c}osa 36570-900, MG, Brazil}
\author{F\'abio D. A. Aar\~ao Reis}
\ead{fdaar@protonmail.com}
\address{Instituto de F\'\i sica, Universidade Federal Fluminense,\\
Avenida Litor\^anea s/n, Niter\'oi, RJ, 24210-340, Brazil}

\begin{abstract}

Using kinetic Monte Carlo simulations, we develop a framework to relate morphological properties
and microscopic dynamics during island growth, coalescence, and initial
formation of continuous heteroepitaxial films.
The average island width is controlled by adatom mobility on the substrate.
Subsequent evolution strongly depends on the Ehrlich-Schw\"{o}ebel energy barrier $E_{ES}$
of the deposited material.
As $E_{ES}$ decreases, islands becomes taller and their coalescence is delayed.
For small islands and large $E_{ES}$, the global roughness increases
as $W\sim{\text{thickness}}^{1/2}$ and the local roughness increases
at short scales (apparent anomalous scaling) before and after island coalescence.
If the islands are wide, $W$ may have a plateau for large $E_{ES}$ and has a maximum
for small $E_{ES}$ when the islands coalesce.
This framework is applied to atomic-force microscopy data of the initial stages of CdTe
deposition on Kapton:
a maximum of $W$ during island coalescence indicates negligible ES barrier,
consistently with scaling properties of much thicker films, and the diffusion coefficient
${10}^{-7}{\text{--}}{10}^{-5}{\text{cm}}^2/{\text s}$ on the Kapton surface at
$150\,^{\circ}\mathrm{C}$ is estimated.
Applications of the framework to other materials are suggested, in which the expected roles
of ES barriers are highlighted.

\end{abstract}


\end{frontmatter}


\section{Introduction}
\label{intro}

Solid films deposited from vapor or in solution have a large variety
of technological applications \citep{ohring},
which explain the efforts to understand the relations
between their morphological properties and microscopic dynamics.
The film properties depend on the initial stages of deposition because
the nucleation, growth, and coalescence of islands affect, for instance, the grain
crystallography and the formation of grain boundaries
\citep{michely,zhangSci1997,ratsch2003,etb}.
Modeling the microscopic interactions of atoms or molecules may be useful to determine those
properties.
For instance, the geometry of islands and mounds in the homoepitaxial growth of some metals
was already described and the activation energies of adatom diffusion 
in different local surface configurations was determined
\citep{stroscio1995,kalff1997,stoldt2000,furman2000,caspersen2002,yuPRL2002}. 
However, this is more difficult when atoms or molecules are deposited on weakly interacting
substrates, in which wide and thick three-dimensional islands are formed (Volmer-Weber mode).
This growth mode is observed in the deposition of a large variety of materials,
such as metals, semiconductors, molecular compounds, and organic molecules,
depending on the substrate used.
For the deposition of thermally evaporated transition metals on substrates such as oxides
and graphene, some atomistic models were also developed; in general, they are nontrivial
extensions of the models used for homoepitaxy
\citep{warrenderPRB2007,elofsson2014,luPRMat2018}.

On the other hand, after a continuous film is formed,
dynamic scaling of the surface fluctuations is frequently used to identify 
the universality class of film roughening \citep{barabasi}.
This approach helps to infer the main physical processes that control the growth
and was already applied to deposits of various materials \citep{barabasi,zhao,krug}.
Several works have also investigated the scaling of surface fluctuations and
correlations in the initial stages of heteroepitaxial film deposition
\citep{zhangSS2007,guo2010,leeJPSJ2011,yangJAP2012,gedda2014,liao2016,liuMatResExp2017,
agrawal2018,creeden2019,pradhan2019,spreitzer2019,chiodini2020,parveen2020,gervilla2019ApplSurfSci}.
Their results cannot be interpreted in the light of kinetic roughening theories
because the deposits are mostly formed by isolated islands, so that height fluctuations
and correlations are related to island widths, heights, and surface density.
Thus, a relevant question is whether surface fluctuations and correlations at those initial
deposition stages can be related to the microscopic growth dynamics,
independently of a specific atomistic modeling and of dynamic scaling analysis.

The first aim of our work is to investigate this question in a model for heteroepitaxial
film growth using kinetic Monte Carlo simulations.
The study considers a broad range of parameters that represent the interplay of atomic 
or molecular flux and their diffusion after adsorption.
The surface roughness (local and global) and the autocorrelation function are measured
from the island growth to the formation of a continuous film.
In the transition between these regimes, those quantities are shown to be related to
microscopic parameters that control the film growth, with
particular relevance for the Ehrlich-Schw\"{o}ebel (ES) \citep{ES} barriers at step edges.
Applications to previous works on growth of metalic and organic films are discussed.

Our second aim is to validate this theoretical framework in the study of the
initial stages of CdTe film deposition on Kapton.
CdTe is a direct gap semiconductor widely used in solar cells and other optoelectronic
devices \citep{major2016} and polyimide substrates are advantageous for their lightness
and flexibility \citep{mathew2004,romeo2018}.
These substrates constrain deposition to relatively low temperatures (compared e.g. with glass),
but this feature is important to reduce production costs \citep{romeo2018}
and CdTe cells on Kapton have conversion efficiencies exceeding $11\%$ \citep{salavei2016}.

Here, the observed features in the transition from CdTe island growth to continuous film
formation suggest the presence of very small ES barriers, a result
consistent with the previous observation of
Kardar-Parisi-Zhang (KPZ) \citep{kpz} scaling of surface fluctuations
in much thicker CdTe films.

\section{Models and Methods}
\label{modelmethod}

\subsection{Experimental Methods}
\label{experimentalmethod}

We thermally evaporated CdTe material ($99.99\%$ purity, Sigma Aldrich) on 
$1.0~{\text{cm}} \times 1.0$~cm polyimide foils (Kapton$^{\textregistered}$ HN, DuPont)
using a home made hot wall epitaxy (HWE) growth system \citep{ferreira2003}. 
The HWE system consists of two independently controlled furnaces, used for source and substrate, 
separated by a shutter that prevents CdTe vapour to impinge on the substrate during
thermalization and controls the beginning and the end of the growth process.
The furnaces are installed inside a vacuum chamber.
The deposition parameters are shown in Table \ref{tabelaexp}.

\begin{table}[!ht]
\caption{Parameters of CdTe deposition.}
\centering
\begin{tabular}{|l|l|}
\hline
Pressure of vacuum chamber              & $\sim 10^{-7}$~torr                 \\ \hline
Reservoir temperature  & $510\,^{\circ}\mathrm{C}$           \\ \hline
Substrate temperature  & $150\,^{\circ}\mathrm{C}$           \\ \hline
Deposition rate        & $14.0\pm0.3$~nm/min                 \\ \hline
Deposition times (min) & $3.0$, $7.5$, $9.0$, $15.0$, $25.0$ \\ \hline
\end{tabular}
\label{tabelaexp}
\end{table}

The substrate temperature used in our system is low compared to most industrial techniques
to produce CdTe cells due to the choice of the flexible Kapton substrate.
However, the high temperature techniques require controlled deposition atmospheres
to avoid the desorption of CdTe from the substrate \citep{romeo2018},
which is not the case here.
In the low temperature vapor deposition, no remarkable difference in CdTe crystal quality
has been observed \citep{romeo2018}.
Moreover, this low temperature technique produces films with smaller roughness,
which can be advantageous for optical applications \citep{ferreiraAPL2006}.

The surface of each film was imaged with atomic force microscopy (AFM) (Ntegra Prima SPM, NT-MDT)
operated in air, contact mode, using Si probes (CSG30, NT-MDT; radius 10 nm)
\citep{almeida2015}.
We imaged several independent square patches of lateral size $L =10~\mu$m over each film surface,
otherwise specified. Lateral and vertical spatial resolutions were near $0.02~\mu$m and $0.1$~nm, 
respectively. 
Each AFM image defines a scalar height field $h(x,y; t_\textrm{d})$,
from which we computed the global and local roughnesses
[Eqs. (\ref{defW}) and (\ref{defw}) in Sec. \ref{basic}]
and the auto-correlation function [Eq. (\ref{defcorr}) in Sec. \ref{basic}].

\subsection{Simulation Methods}
\label{theoretical}

\subsubsection{Deposition Model}
\label{modeldeposition}

The model is defined in a simple cubic lattice in which the lattice constant is the length unit
and the lateral size is $L$.
The initially flat substrate is located at $z=0$, the deposit grows in $z>0$, and
periodic boundary conditions are considered in the $x$ and $y$ directions.
Each atom or molecule of the deposit occupies a single lattice site;
for simplicity, hereafter we refer to the deposited species as an adatom.
Solid-on-solid conditions are considered, so the films have no pores.
A column of the deposit is defined as the set of adatoms with the same $\left( x,y\right)$ position;
the height variable $h\left( x,y\right)$ is the value of $z$ of the topmost adatom in that column.

The deposition occurs with a collimated flux of $F$ atoms per substrate site per unit time.
In each deposition event, a column $\left( x,y\right)$ is randomly chosen and the adsorption of
the incident atom occurs as it lands at the top of that column.
Desorption is neglected.

Surface diffusion is modeled by hops of the adatoms at the top of the columns,
with rates depending on their number of neareast neighbors (NNs).
Substrate atoms/molecules are immobile.
For a simple description of heteroepitaxy, the diffusion coefficient of an adatom
is different at $z=1$, where it is in contact with the substrate,
and at the other layers, $z\geq 2$, where it is in contact only with the same species.
If the adatom has no lateral NN (i.e. if it is on a terrace), the
numbers of hop attempts per unit time depends on its current height:
$D_S$ if it is at $z=1$, $D_A$ if it is at $z\geq 2$.
These rates have activated Ahrrenius forms
\begin{equation}
D_S=\nu\exp{\left[ -E_S/\left(k_BT\right)\right]} \qquad ,\qquad 
D_A=\nu\exp{\left[ -E_A/\left(k_BT\right)\right]} ,
\label{defDSDA}
\end{equation}
where $\nu$ is a frequency, $E_S>0$ and $E_A>0$ are activation energies,
$k_B$ is the Boltzmann constant, and $T$ is the substrate temperature.
Otherwise, if the adatom has $n$ lateral NNs, the hopping rates $D_0$ at $z=1$
and $D_1$ at $z\geq 2$ are 
\begin{equation}
D_0=D_S\epsilon^n \qquad , \qquad D_1=D_A\epsilon^n ,
\label{defD0DA}
\end{equation}
where
\begin{equation}
\epsilon \equiv \exp{\left[ -E_B/\left(k_BT\right)\right]} .
\label{defepsilon}
\end{equation}
Here, $E_B>0$ is a bond energy that represents the interaction with each lateral NN.

\begin{figure}[!h]
\center
\subfloat[]{\includegraphics[width=0.27\textwidth]{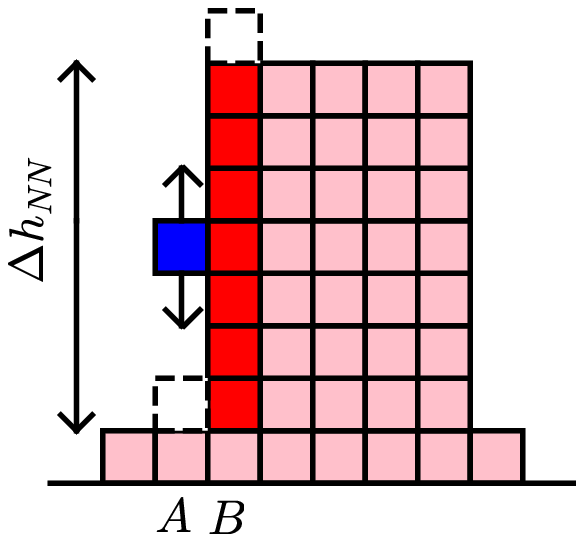}} 
\subfloat[]{\includegraphics[width=0.5\textwidth]{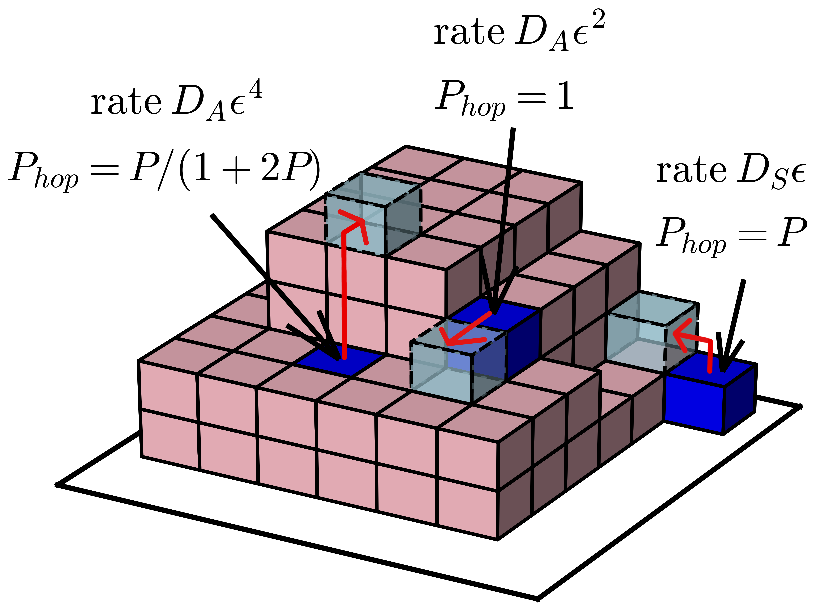}}
\caption{
(a) Two-dimensional view of the processes involved in the hop of an adatom (blue square)
between the top positions (dashed squares) of columns A and B.
The hop represents the adatom diffusion along a vertical wall (red squares) and the crossing of an
edge near the top position.
Other adatoms are shown in pink color.
(b) Hopping rates of three adatoms (blue cubes) and probabilities of 
hop attempts (red arrows) to NN columns (positions indicated by translucent gray cubes).
}
\label{hops}
\end{figure}

When an adatom attempts to hop, the direction is randomly chosen among the four
NN columns, $\pm x$ or $\pm y$.
If the chosen column has the same height (i.e. a hop in the same layer),
the adatom moves with probability $1$.
Otherwise, if the adatom attempts to hop to a layer at a different height $z$, there is
an additional energy barrier for crossing the step edge, which is the Ehrlich-Schw\"{o}ebel (ES)
barrier \citep{ES}, with activation energy $E_{ES}$.
If the height difference ${\Delta h}_{NN}$ between the columns is large [e.g. columns A and B of
Fig. \ref{hops}(a)], the solid-on-solid constraint in a simple cubic lattice
implies that the adatom has to hop from the top of one column to the top of the other.
Thus, besides representing the step edge barrier, the probability of the hop between those
points also has to represent the adatom motion along a vertical terrace of length ${\Delta h}_{NN}$
separating the initial and final positions; see Fig. \ref{hops}(a).
The relevance of mass transport in faceted island walls is shown, for instance,
in recent simulations of Ag deposition on weakly interacting substrates \citep{luPRMat2018}.
Considering that the probability of the adatom to return to the original position
before reaching the other layer increases with ${\Delta h}_{NN}$,
Ref. \citep{lealJPCM} proposed that the overall probability of executing the hop
should be expressed as
\begin{equation}
P_{hop}=\frac{P}{1+P\left({\Delta h}_{NN}-1\right)} \qquad , \qquad
P=\exp{\left[ -E_{ES}/\left(k_BT\right)\right]} .
\label{defP}
\end{equation}
With probability $1-P_{hop}$, the adatom remains at the current position.
Also note that the probability $P_{hop}$ is considered for downwards and upwards hops.

Fig. \ref{hops}(b) illustrates the possible hops of some adatoms and show their
corresponding probabilities.

The interplay between temperature and flux is represented by the
diffusion-to-deposition ratios defined as
\begin{equation}
R_S \equiv \frac{D_S}{F} = \frac{\nu}{F} \exp{\left[ -E_S/\left(k_BT\right)\right]} \qquad ,\qquad 
R_A \equiv \frac{D_A}{F} = \frac{\nu}{F} \exp{\left[ -E_A/\left(k_BT\right)\right]} .
\label{defR}
\end{equation}
They are interpreted as the average numbers of hops of terrace adatoms during the
average time $1/F$ of deposition of one atomic layer.
In heteroepitaxy, the adatom interaction with the substrate is weaker
than the interaction with other adatoms, so that $E_S<E_A$; consequently,
$D_S>D_A$ and $R_S>R_A$.
Throughout this work, the dimensionless parameters that represent the growth conditions
are the ratios $R_S$ and $R_A$, the detachment probability $\epsilon$, and
the probability $P$ of crossing a monolayer step edge.

This model is an extension of the Clarke-Vvedensky (CV) model \citep{cv}
of thin film growth, in which the hop frequency is $\nu=k_BT/\left(\pi\hbar\right)$,
where $\hbar$ is the reduced Planck's constant, as predicted by transition state theory.
However, it is also frequent that a constant $\nu$ (set e.g. as ${10}^{12}{\text s}^{-1}$) is used
in simulation and analytical works \citep{etb}.
This choice weakly affects the results because temperature variations have much stronger effects
on the Boltzmann factors of Eqs. (\ref{defDSDA}), (\ref{defepsilon}),  and (\ref{defP}).

In the presentation of simulation results, we use the dimensionless film thickness
\begin{equation}
d=Ft .
\label{defd}
\end{equation}
This is the average number of deposited atoms per substrate site at time $t$ or,
alternatively, the number of deposited monolayers.

\subsubsection{Simulation Details}
\label{simulationmethod}

We performed simulations in lattices with $L=1024$, in which finite size effects are small.
The maximal dimensionless film thickness is $d=640$.
For each parameter set, we generated $10$ deposits and observed small fluctuations in
the average quantities.

Table \ref{tabelaparametros} shows the simulated parameter sets that are presented in this work,
labeled from A to P.
Simulations with other parameter sets were also performed, which support the conclusions stated here.
The ratio $R_A$ ranges between ${10}^3$ and ${10}^6$, which is typical of epitaxial growth
in broad temperature ranges \citep{etb,toreis2020}.
The ratio $R_S$ is ${10}^{2}$--${10}^{3}$ times larger than $R_A$, which corresponds
to a facilitated diffusion in the substrate.
The detachment probability $\epsilon$ ranges between ${10}^{-4}$ and ${10}^{-2}$.
The step-edge probability $P$ is chosen in the range ${10}^{-3}$--$1$, where the highest value
indicates the absence of the ES barrier.

\begin{table}[!ht]
\caption{Sets of dimensionless parameters used in the simulations.}
\centering
\begin{tabular}{|l|l|l|l|l|}
\hline
Set & $R_S$    & $R_A$    & $\epsilon$  & $P$    \\ \hline
A   & ${10}^6$ & ${10}^4$ & ${10}^{-3}$ & $0.01$ \\ \hline
B   & ${10}^6$ & ${10}^4$ & ${10}^{-3}$ & $0.1$  \\ \hline
C   & ${10}^6$ & ${10}^4$ & ${10}^{-3}$ & $1$    \\ \hline
D   & ${10}^7$ & ${10}^5$ & ${10}^{-3}$ & $0.01$ \\ \hline
E   & ${10}^7$ & ${10}^5$ & ${10}^{-3}$ & $0.1$  \\ \hline
F   & ${10}^7$ & ${10}^5$ & ${10}^{-3}$ & $1$    \\ \hline
G   & ${10}^7$ & ${10}^5$ & ${10}^{-2}$ & $0.01$ \\ \hline
H   & ${10}^7$ & ${10}^5$ & ${10}^{-2}$ & $0.1$  \\ \hline
I   & ${10}^7$ & ${10}^5$ & ${10}^{-2}$ & $1$    \\ \hline
J   & ${10}^8$ & ${10}^5$ & ${10}^{-3}$ & $0.01$ \\ \hline
K   & ${10}^8$ & ${10}^6$ & ${10}^{-2}$ & $0.01$ \\ \hline
L   & ${10}^8$ & ${10}^6$ & ${10}^{-2}$ & $0.1$  \\ \hline
M   & ${10}^8$ & ${10}^6$ & ${10}^{-2}$ & $1$    \\ \hline
N   & ${10}^9$ & ${10}^6$ & ${10}^{-2}$ & $0.01$ \\ \hline
O   & ${10}^9$ & ${10}^6$ & ${10}^{-2}$ & $0.1$  \\ \hline
P   & ${10}^9$ & ${10}^6$ & ${10}^{-2}$ & $1$    \\ \hline
\end{tabular}
\label{tabelaparametros}
\end{table}

The simulations are implemented with a kinetic Monte Carlo algorithm
described in detail in Ref. \citep{toreis2020}.

\subsection{Basic Quantities}
\label{basic}

The height fluctuations are monitored with the calculation of the global
roughness $W$ defined as
\begin{equation}
W\equiv
\left< {\left[~ \overline{{\left( h - \overline{h}\right) }^2}  ~\right] }^{1/2} \right>  .
\label{defW}
\end{equation}
Here the overbars denote a spatial average and the angular brackets denote
an average over different configurations with a given thickness.
The asperity of the surface in square observation windows of lateral size $r$
is quantified by the local surface roughness
\begin{equation}
w\equiv \left< {{\left<{{\left( h-{\left<h\right>}_r\right)}^2}\right>}_r}^{1/2}\right> ,
\label{defw}
\end{equation}
where the inner brackets (with subindex $r$) denote the averages inside an observation window
and the outer brackets denote averages over different window positions and over different deposits
with the same thickness.

The autocorrelation function, which measures the correlations in 
the heigth fluctuations at a distance $s$, is defined as \citep{zhao}
\begin{equation}
\Gamma \equiv \frac{ \left\langle {\left[ \tilde{h}\left( {\vec{r}}_0+\vec{s}\right)
-\tilde{h}\left( {\vec{r}}_0\right) \right]}^2 \right\rangle }{W_2} \qquad , \qquad
s\equiv |\vec{s}| \qquad , \qquad \tilde{h}\equiv h-\overline{h} \qquad .
\label{defcorr}
\end{equation}
Here the configurational average is taken over different initial positions
${\vec{r}}_0$, different orientations of $\vec{s}$ (directions $x$ and $y$), and different deposits,
and $W_2$ is a global square roughness defined as in Eq. (\ref{defW}) but without taking
the square root of the spatial average.
With this definition, $\Gamma=1$ for $s=0$ in any thickness.

The autocorrelation function of mounded surfaces usually has the first minimum at 
a position $s_M$ that approximates the average distance
between the top of the hills and the bottom of the valleys \citep{zhao}.
Here, $s_M$ is used to characterize the deposit patterns before, during, and after
island coalescence.

In the simulation data, $W$, $w$, and $s_M$ are dimensionless quantities measured in
units of the lattice constant.
In the experimental data, the global and the local roughnesses
are in nanometers, whereas $s_M$ is in micrometers.
The height $h$ is measured in nanometers, as explained in Sec. \ref{experimentalmethod}.

During the simulations, isolated islands are formed at short times and eventually coalesce.
Our definition of islands considers only the occupation of the level $z=1$, in which two adatoms
belong to the same island if they are NNs.
The enumeration of islands is performed with the Hoshen-Kopelman algorithm \citep{hoshen}.
The island density $N_{isl}$ is defined as the number of islands per substrate site.

\section{Results of $\text{CdTe}$ Deposition}
\label{resultCdTe}

\begin{figure}[!h]
\center
    \includegraphics[width=\textwidth]{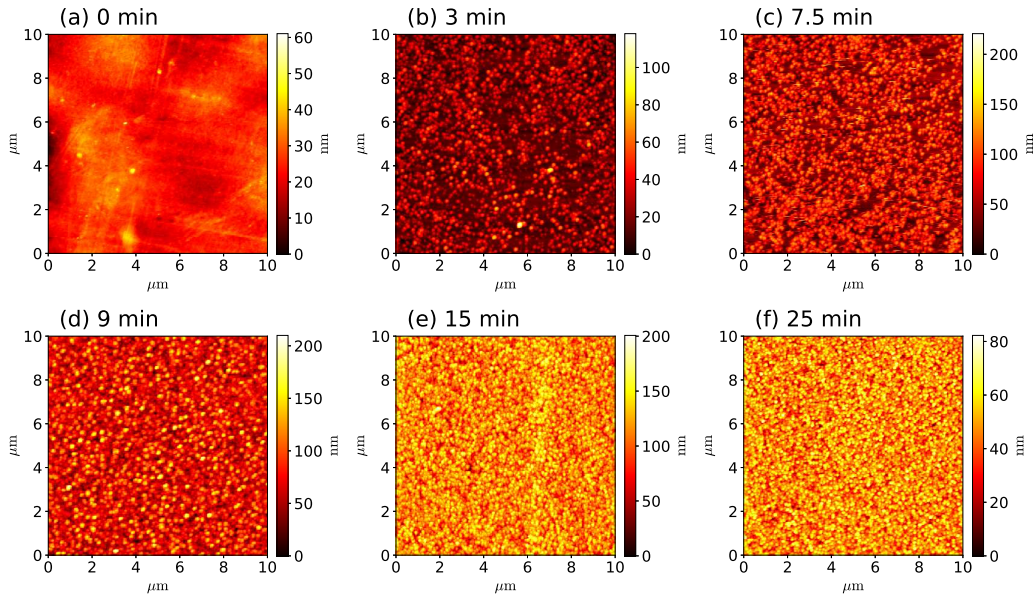}
\caption{
Representative AFM images of (a) the Kapton$^{\textregistered}$ substrate surface and
of (b)-(f) the CdTe deposits at the times indicated.
}
\label{figCdTe}
\end{figure}

Figures \ref{figCdTe}(a)-(f) show typical surface images of the initial stages of CdTe deposition.
The substrate surface [Fig. \ref{figCdTe}(a), $t=0$] has $W\approx5$~nm.
At $t=3$~min, several islands with heights up to $100$~nm are formed
and the average thickness is $42\pm1$~nm.
The deepest regions in Fig. \ref{figCdTe}(b) (darkest colors) have CdTe coverages
much smaller than the average thickness or the island heights.
At $t=7.5$~min, the average thickness is $105\pm2$~nm and the heights of several islands are
close to $200$~nm, so the same interpretation holds for the deepest regions
of the image in Fig. \ref{figCdTe}(c).
At $t=15$~min [Fig. \ref{figCdTe}(e)],
a different scenario starts to appear because the maximal height differences
and the average thickness have approximately the same value $200$~nm, which suggests that
most of the substrate area (possibly all that area) is already covered by CdTe.
In other words, at that time, island coalescence is complete or almost complete.
At the longest time [$t=25$~min; Fig. \ref{figCdTe}(f)], the average thickness increased
to $350$~nm but the height fluctuations had a sharp decrease and are limited to $80$~nm.
This indicates that a smoothening process takes place in the transition from island coalescence
to continuous film formation.

\begin{figure}[!h]
\center
    \includegraphics[width=\textwidth]{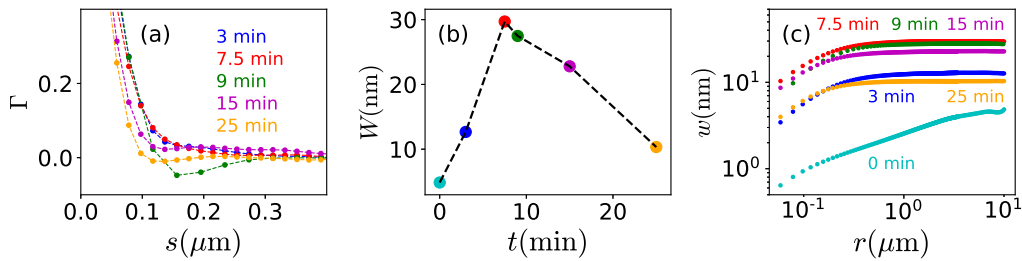}
\caption{
(a) Autocorrelation function, (b) global roughness, and (c) local roughness
of CdTe deposits at several growth times:
$t=0$ (cyan), $3$~min (blue), $7.5$~min (red), $9$~min (green), $15$~min (magenta),
and $25$~min (orange).
Dashed lines are drawn to guide the eye.
}
\label{dadosCdTe}
\end{figure}

Figure \ref{dadosCdTe}(a) shows the autocorrelation function at the same deposition times.
The curves for $t\geq9$~min show minima at $s\approx0.1{\text{--}}0.15~\mu$m, which are of
the same order of magnitude as the island sizes.
Figure \ref{dadosCdTe}(b) shows the evolution of the global roughness.
It has the largest values $W\approx30$~nm at $t=7.5$~min and $9$~min,
and then decreases to a value $\approx10$~nm at $t=25$~min.
In Ref. \citep{almeida2017}, it was shown that the roughness of these
CdTe films fluctuates between $10$~nm and $20$~nm until $\approx200$~s of deposition.
This confirms that there is smoothening of the surface after the
island coalescence and the initial formation of a continuous film.

Figure \ref{dadosCdTe}(c) shows the local roughness $w$ as a function of the window size $r$
for several growth times.
The plots at $t=3$~min and $t=7.5$~min show $w$ increasing
in time for all $r$.
This split of a $w\times r$ plot in all scales is a typical feature of systems with
anomalous roughening \citep{ramasco}.
However, the interpretation of the present results cannot follow theories of
anomalous roughening  because this is a transient feature related to the 
initial island pattern.
Similar conclusion was drawn in Ref. \citep{almeida2015}.
From $t=7.5$~min to $15$~min, the changes in $w$ for all window sizes are smaller than
the changes formerly observed.
Finally, at $t=25$~min, the roughness has decreased at all scales,
confirming the smoothening when the continuous film is formed.

\section{Simulation Results}
\label{simulationresults}

\subsection{Island Growth}
\label{islandgrowth}

Here we analyze the island features for deposition of one atomic layer,
i.e. dimensionless film thickness $d=1$.

\begin{figure}[!h]
\center
    \subfloat[]{\includegraphics[width=0.22\textwidth]{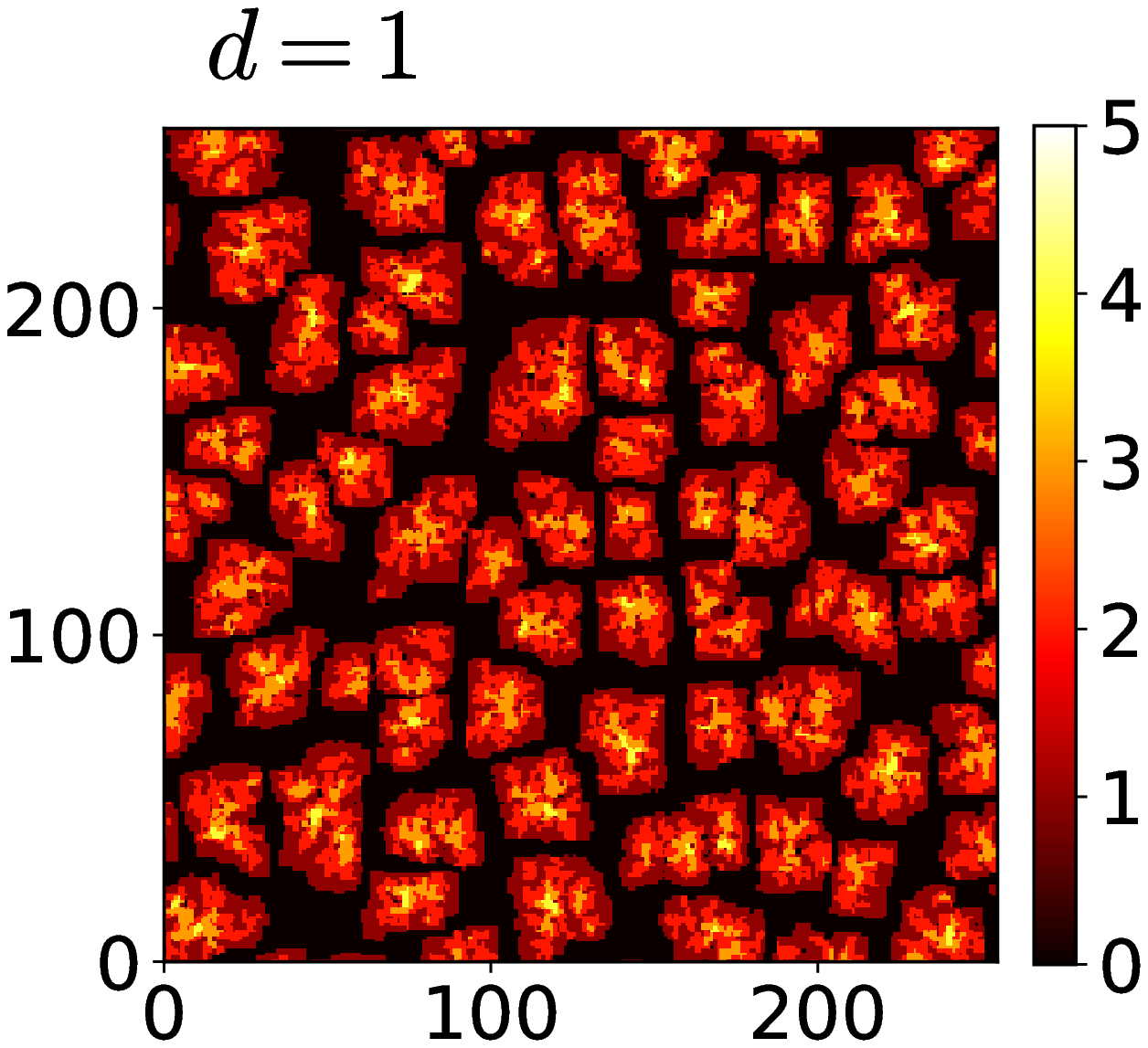}}\hskip 3mm
    \subfloat[]{\includegraphics[width=0.22\textwidth]{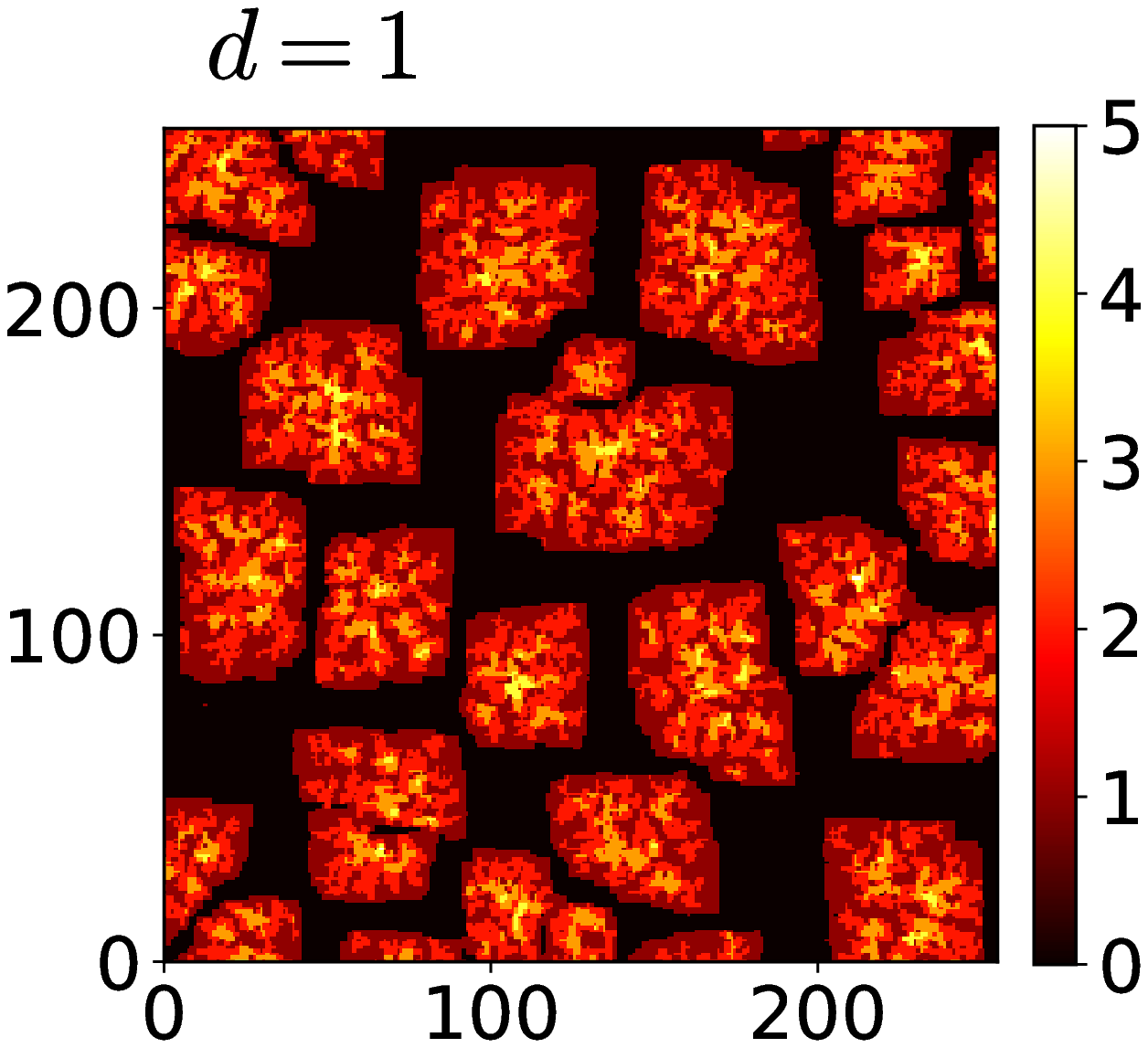}}\hskip 3mm
    \subfloat[]{\includegraphics[width=0.22\textwidth]{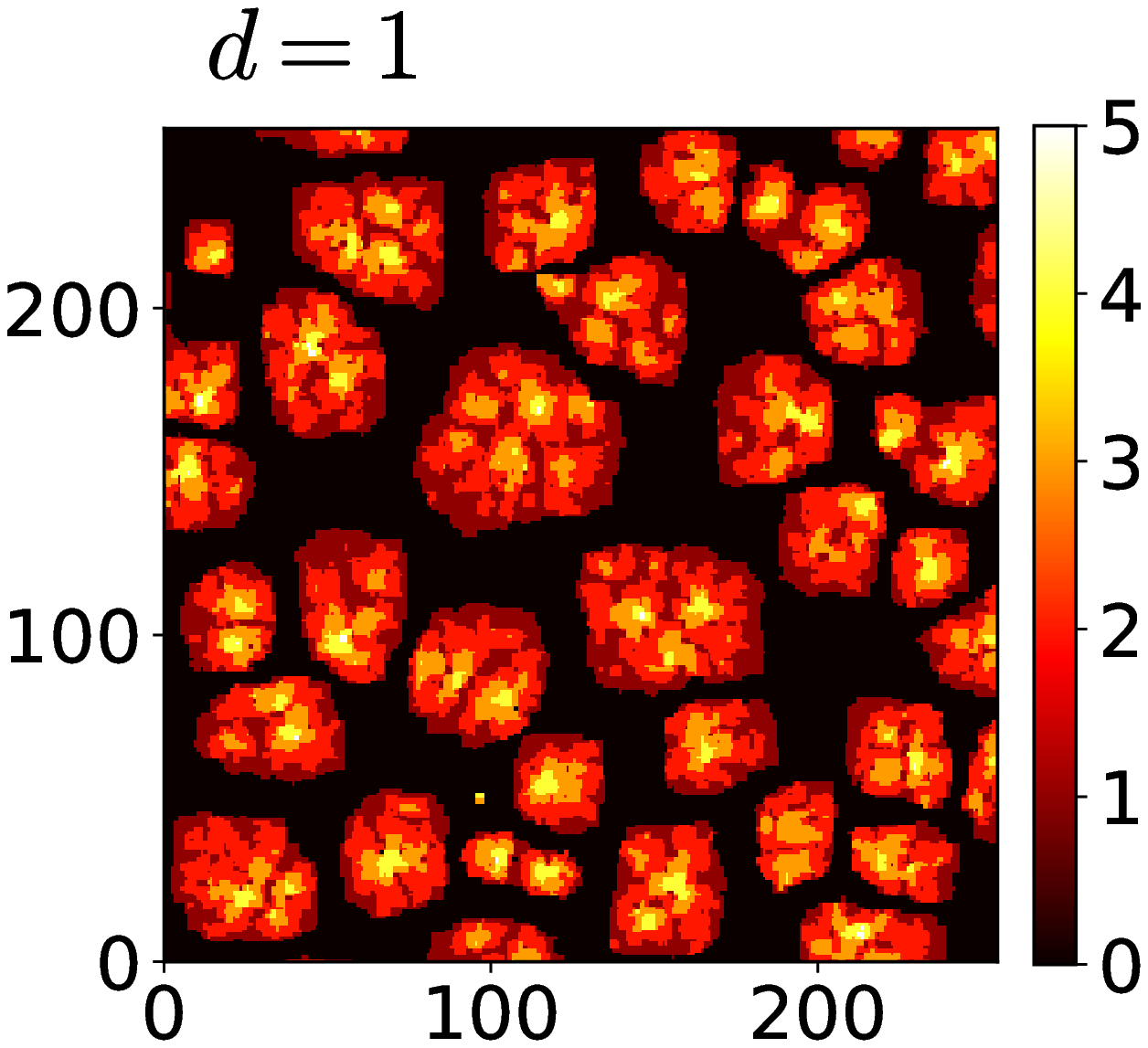}}\hskip 3mm
    \subfloat[]{\includegraphics[width=0.22\textwidth]{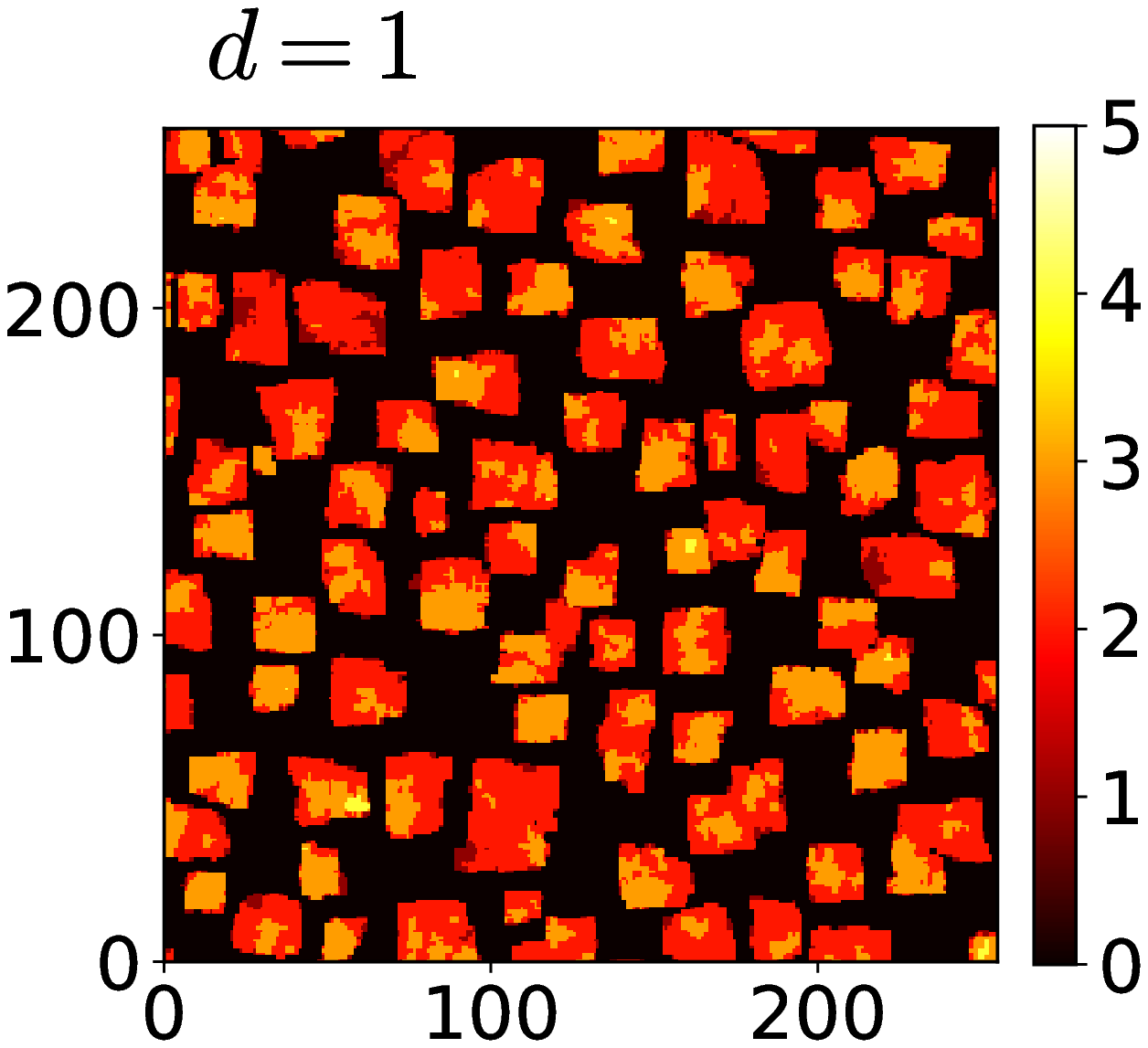}}
\caption{
Top views of parts of the deposits produced in simulations with average thickness of one layer.
The parameter sets are (a) D, (b) J, (c) G, and (d) F.
All lengths are in units of the lattice constant.
}
\label{monolayer}
\end{figure}

Figure \ref{monolayer}(a) shows an image of a deposit obtained with
the parameter set D, in which the bonds with lateral NN are strong and the ES barrier is large.
The average island size is small and their borders are disordered.
At $z\geq 2$, branched clusters are formed because $R_A$ and $\epsilon$ are relatively small
and prevent the adatom detachment from NNs.
A small occupation of the layers with $z\geq 3$ is observed due to restrictions to interlayer
transport.
If $R_S$ or $\epsilon$ are smaller than the present values by two or more orders of magnitude,
branched islands are also formed in the first layer.

Figure \ref{monolayer}(b) shows an image of a deposit obtained with a larger diffusion
coefficient at the substrate in comparison with Fig. \ref{monolayer}(a).
The island size increases because coarsening is facilitated during the initial deposition stages,
but branched structures are still being formed on the top of the islands.
We checked that changes in $R_A$ have little effect on island density (measured at $z=1$),
but larger $R_A$ produces more compact structures at the layers  with $z\geq2$.
Figure \ref{monolayer}(c) shows an island configuration obtained with the same parameters as
in Fig. \ref{monolayer}(a), but with larger detachment probability $\epsilon$. 
This leads to a decrease in island density (increase in the island size) and
helps the formation of more compact structures at all layers.

Finally, Fig. \ref{monolayer}(d) shows the effect of a negligible step-edge barrier, $P=1$,
while the other parameters remain the same as in Fig. \ref{monolayer}(a).
This change has little effect on island density.
However, now the islands are higher and more compact;
the clusters at the upper layers ($z\geq 2$) are also more compact.
This is a consequence of facilitated transport from the
substrate to the top of the islands, particularly for the adatoms with
small numbers of lateral bonds.
The widths of the gaps between the islands increase, so
a smaller ES barrier delays island coalescence (as will be confirmed later).

Previous works on submonolayer island growth, in which the dynamics is restricted to the
first layer, obtained scaling relations for the island density $N_{isl}$ in terms of
$R_S$ and $\epsilon$ for small coverages (typically $d\lesssim0.1$)
\citep{ratsch1995,barteltSSL1995,submonorev}.
In conditions of facile detachment of low-bonded atoms ($n=1$) from islands,
it was predicted that $\epsilon R_S^{2/3}\gtrsim1$ and that the island density scales as
\citep{submonorev}
\begin{equation}
N_{isl}=\epsilon^3 f\left( Y\right) , Y\equiv \epsilon R_S^{1/5} ,
\label{Nisl}
\end{equation}
where $f$ is a scaling function.
For $Y\lesssim1$, the dependence of $N_{isl}$ on $\epsilon$ is weak, so Eq. (\ref{Nisl})
implies $N_{isl}\sim R_S^{-3/5}$ \citep{submonorev}.

\begin{figure}[!h]
\center
\includegraphics[width=0.45\textwidth]{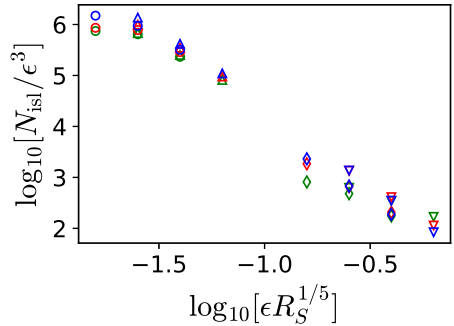}
\caption{
Scaled island densities at dimensionless thickness $d=1$ obtained in simulations.
The parameter sets have:
$R_{S}/R_{A}=10^2$, $\epsilon=10^{-3}$ (circles); 
$R_{S}/R_{A}=10^2$, $\epsilon=10^{-2}$ (diamonds);
$R_{S}/R_{A}=10^3$, $\epsilon=10^{-3}$ (triangles);
$R_{S}/R_{A}=10^3$, $\epsilon=10^{-2}$ (inverted triangles).
The color code distinguishes the ES barrier: $P=1$ (blue); $P=0.1$ (red); $P=0.01$ (green).
}
\label{scalingNisl}
\end{figure}

Here we measured $N_{isl}$ by the occupancy of the level $z=1$ in
deposits with one monolayer ($d=1$).
In this case, there is significant occupancy of the levels $z\geq2$, which differs from
the submonolayer studies mentioned above.
Despite this difference, Fig. \ref{scalingNisl} shows a good data collapse of
$N_{isl}/\epsilon^3$ as a function of $Y$.
The deviations increase as $Y$ increases because the effects of $P$
become more relevant when the mobility on the substrate is very large;
indeed, interlayer transport depends not only on $P$, but also on $R_S$ and $\epsilon$.

These results show that the relations obtained in low coverage submonolayers are also reasonable
approximations for the island density with coverage of one monolayer.
The smallest islands, which are obtained for the smallest $Y$,
have less than ${10}^3$ adatoms in the first layer and widths $\lesssim30$
lattice constants when they begin to coalesce.
This width sets the crossover between the regimes of small and large islands in this model.

\subsection{Film Formation with Large ES Barrier}
\label{largeES}

\begin{figure}[!h]
\center
    \includegraphics[width=\textwidth]{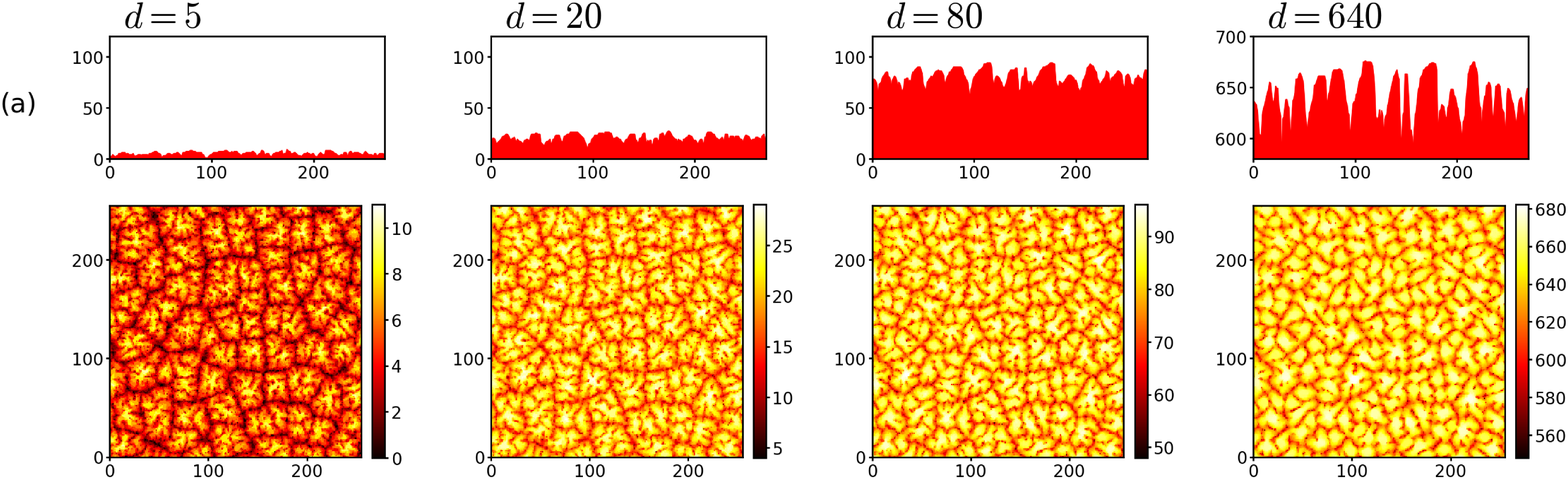} \\
    \vspace{0.5cm}
    \includegraphics[width=\textwidth]{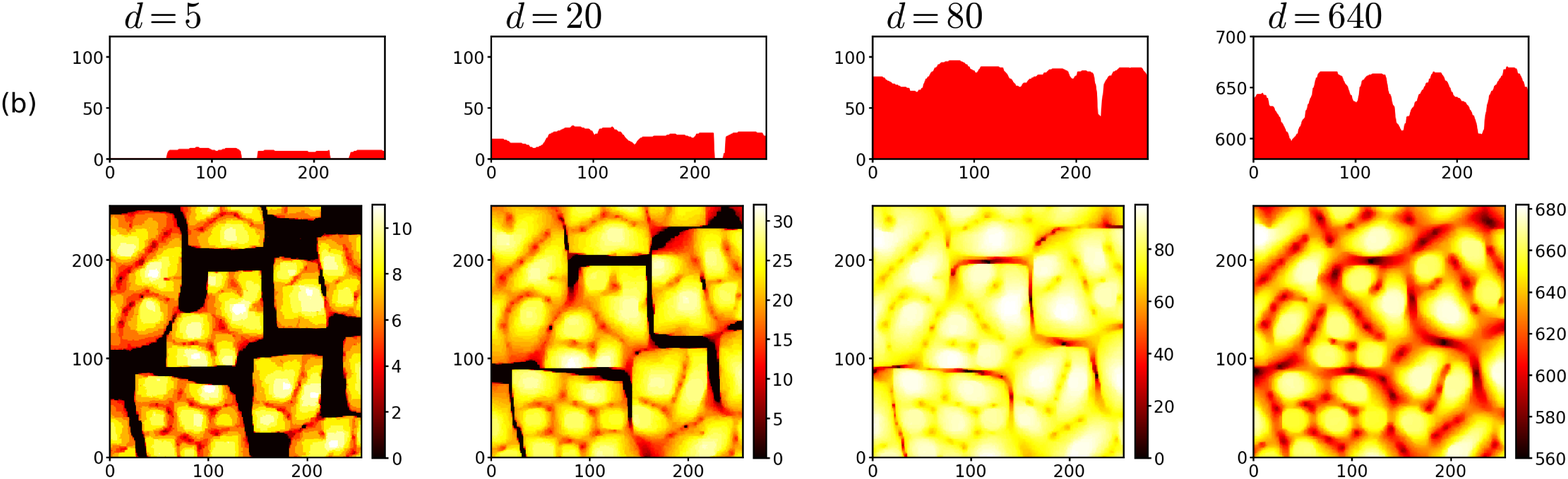}
\caption{
Vertical cross sections and top views of parts of the deposits grown in simulations
with large ES barriers.
Parameter sets are (a) D and (b) K.
Cross sections are not in scale and some of them are restricted to the upper layers.
All lengths are in units of the lattice constant.
}
\label{figPsmall}
\end{figure}

Here we consider $P=0.01$, corresponding to a large ES barrier.
Fig. \ref{figPsmall}(a) and \ref{figPsmall}(b) show top views and cross sections of deposits
with several thicknesses grown with two parameter sets.

In Fig. \ref{figPsmall}(a), the parameters are the same of Fig. \ref{monolayer}(a).
For dimensionless thickness $d=5$, all islands already coalesced into a single large domain,
and for $d=20$ a continuous film is formed (minimum height is $z=5$).
However, until the largest thickness ($d=640$), the gaps between the initial islands
are visible, showing that coarsening is very slow.
Moreover, close inspection shows the formation of smaller sharp structures separated by other gaps.
These features are representative of other deposits with small initial islands, which are
obtained with small $R_S$ and $\epsilon$.

In Fig. \ref{figPsmall}(b), higher adatom mobility is considered.
Larger islands are formed and they have already coalesced when $d=20$,
but some unfilled gaps remain until $d=80$.
It shows that the increase of adatom mobility delays island coalescence.
At the top of each initial island, we observe the formation of some mounds;
the mound widths are smaller than the island sizes because $R_A\ll R_S$ and they
are rounded and smoother than the structures formed in Fig. \ref{figPsmall}(a)
because $R_A$ and $\epsilon$ are larger.
The pattern observed at the longest times mixes the initial island pattern
with the mounded pattern formed on the top of those islands.

\begin{figure}[!h]
\center
\includegraphics[height=0.21\textwidth]{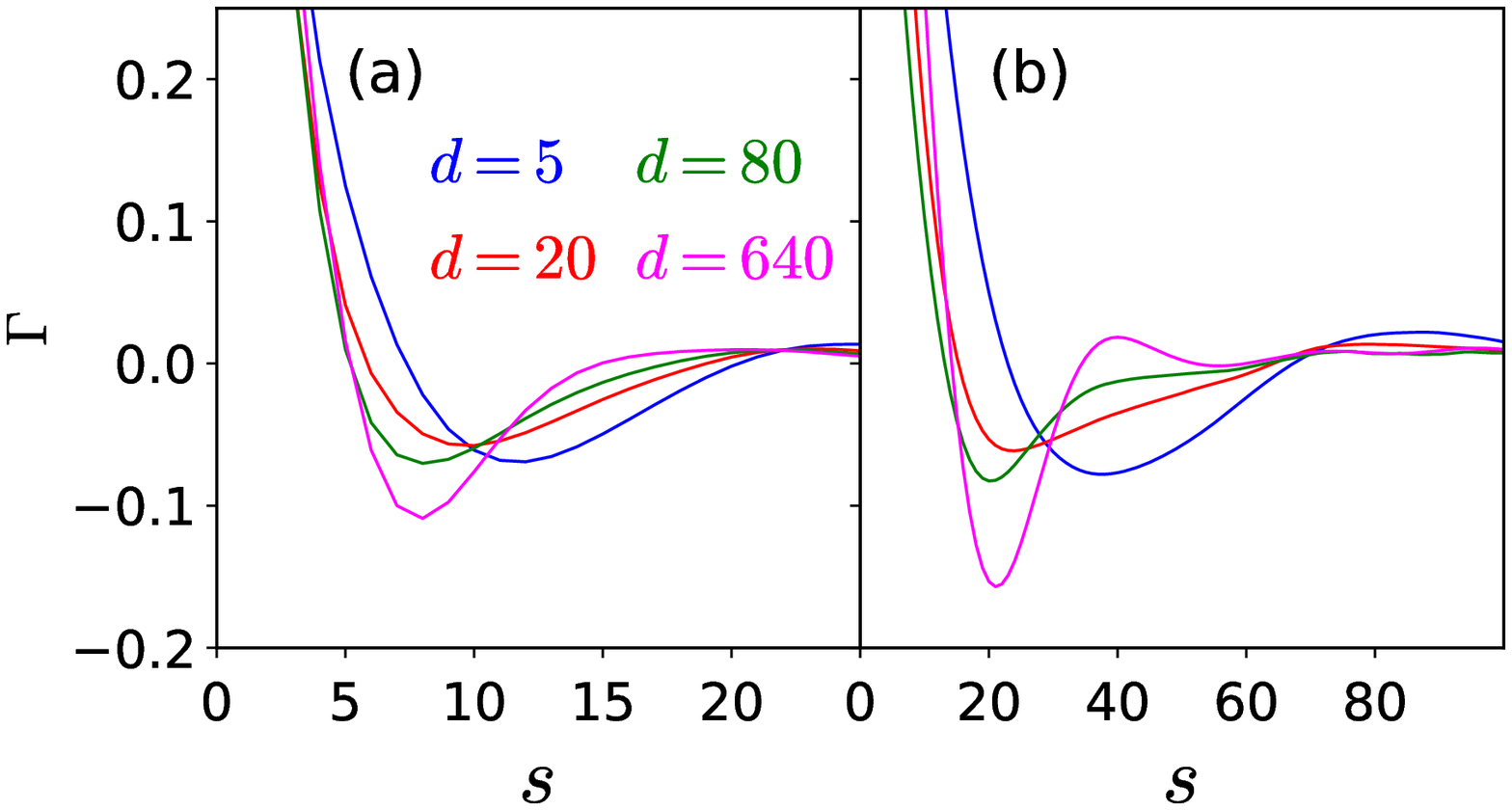}
\includegraphics[height=0.21\textwidth]{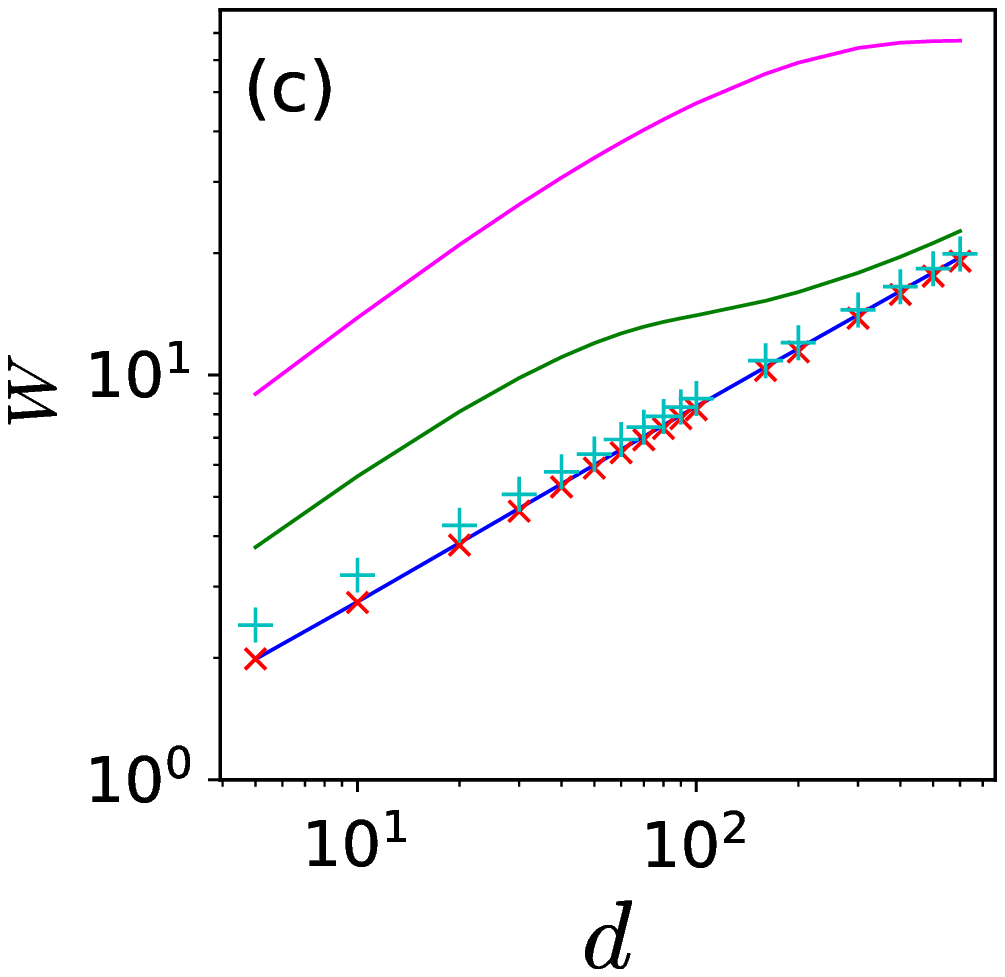}
\includegraphics[height=0.21\textwidth]{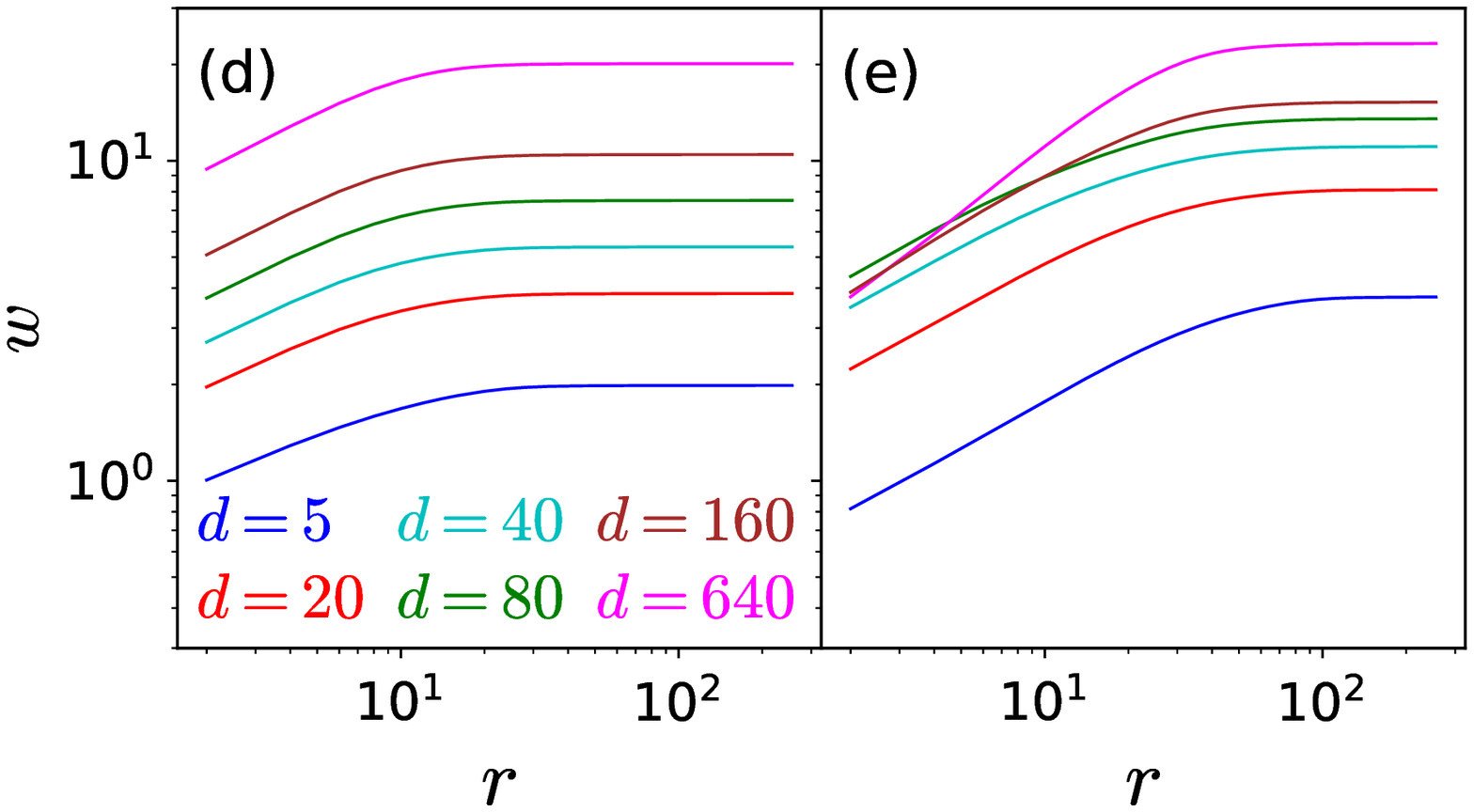}
\caption{
Auto-correlation function [(a),(b)], global roughness [(c)], and local roughness [(d)-(e)]
of films grown in simulations with large ES barrier.
The parameter sets are:  (a),(d): D; (b),(e): K;
(c): A (red crosses), D (blue line), G (cyan plus signs), K (green line), and N (magenta line).
All lengths are in units of the lattice constant.
}
\label{P001}
\end{figure}

Figures \ref{P001}(a) and \ref{P001}(b) show the auto-correlation function $\Gamma$
as a function of the distance $s$ for the same parameters and film thicknesses
of Figs. \ref{figPsmall}(a) and \ref{figPsmall}(b), respectively.
The position of the first minimum of $\Gamma$, which we denote as $s_M$, indicates the typical distance
between the top of the hills and the bottom of the valleys of the film surface.
Before island coalescence and without mound formation on them,
$s_M$ is related to the island size, as observed for $d=5$ in both cases.
As the thickness increases, $s_M$ decreases.
For $d\geq80$, $s_M$ is related to the size of the mounds formed on the top of the initial
islands; the mound slopes are increasing in the continuous film, which explains the deepening
of the $\Gamma$ minimum.

Figure \ref{P001}(c) shows the evolution of the global roughness $W$ for five parameter sets.
In the cases of small islands and rapid coalescence ($d\lesssim5$),
the data collapse into a single universal line with slope
$\approx0.5$; this is the case, for instance, with the parameter set D of the images in
Fig. \ref{figPsmall}(a).
They are cases with effectively random (uncorrelated) deposition, in which adatom diffusion
is mostly restricted to terraces.
However, for $R_S\geq{10}^8$ and $\epsilon\geq{10}^{-2}$, the large adatom mobility on the substrate
facilitates mass transport to the top of the islands, despite the ES barrier.
This leads to larger height differences between those islands and the substrate,
i.e. larger $W$, but $W$ slows down after island coalescence.
For the parameter set K of the images in Fig. \ref{figPsmall}(b),
the $\log{W}\times\log{d}$ plot has an inflection and then continues to increase.

Figures \ref{P001}(d) and \ref{P001}(e) show the local roughness $w$ as a function of 
window size $r$ for the same parameter sets of Figs. \ref{figPsmall}(a) and
\ref{figPsmall}(b), respectively, and several thicknesses.
In both cases, during island growth, an apparent anomalous roughening
is observed, in which $w$ increases for small $r$ as fast as $W$ does (large $r$).
However, this is a transient behavior that should not be interpreted as true anomalous
roughening \citep{ramasco}.
Indeed, the subsequent evolution differs in the two cases.
In Fig. \ref{P001}(d), which is typical for small islands and rapid coalescence,
the anomaly remains after the continuous film is formed.
There are small correlations at short scales, which explain the increase of $w$ for
$r\lesssim 20$, but uncorrelated growth in large scales.
In Fig. \ref{P001}(e), with larger $R_S$, $R_A$, and $\epsilon$, larger islands
are formed and slower variations of $w$ are observed after the continuous film formation,
in small or large windows.

In all cases, when the film thickness is sufficiently large and the memory
of the island pattern has disappeared, the large ES barrier is expected to lead to
true anomalous roughening in all cases, as observed in other deposition processes
with constrained interlayer transport \citep{lealJPCM,lealJSTAT,petrov2014}.
However, the larger the initial islands, the longer the time for attaining this regime.
For this reason, for $R_S={10}^9$, Fig. \ref{P001}(c) only shows a maximum of $W$
for the largest simulated thicknesses.

\subsection{Film Formation without ES Barrier}
\label{zeroES}

\begin{figure}[!h]
\center
     \includegraphics[width=\textwidth]{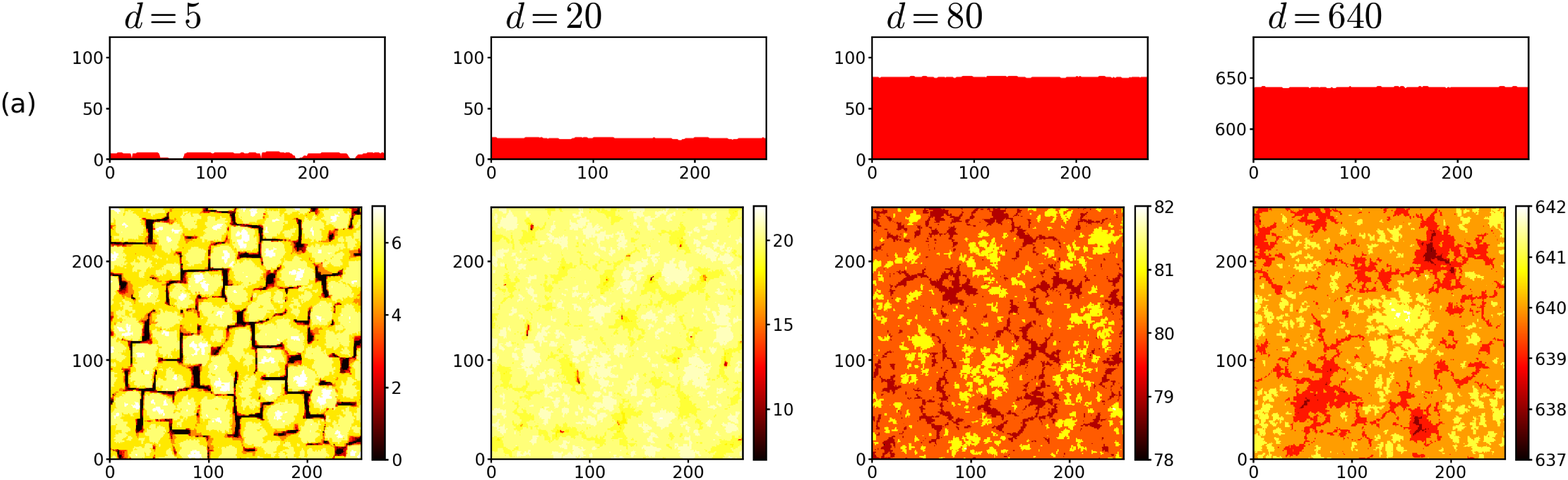} \\
    \vspace{0.5cm}
    \includegraphics[width=\textwidth]{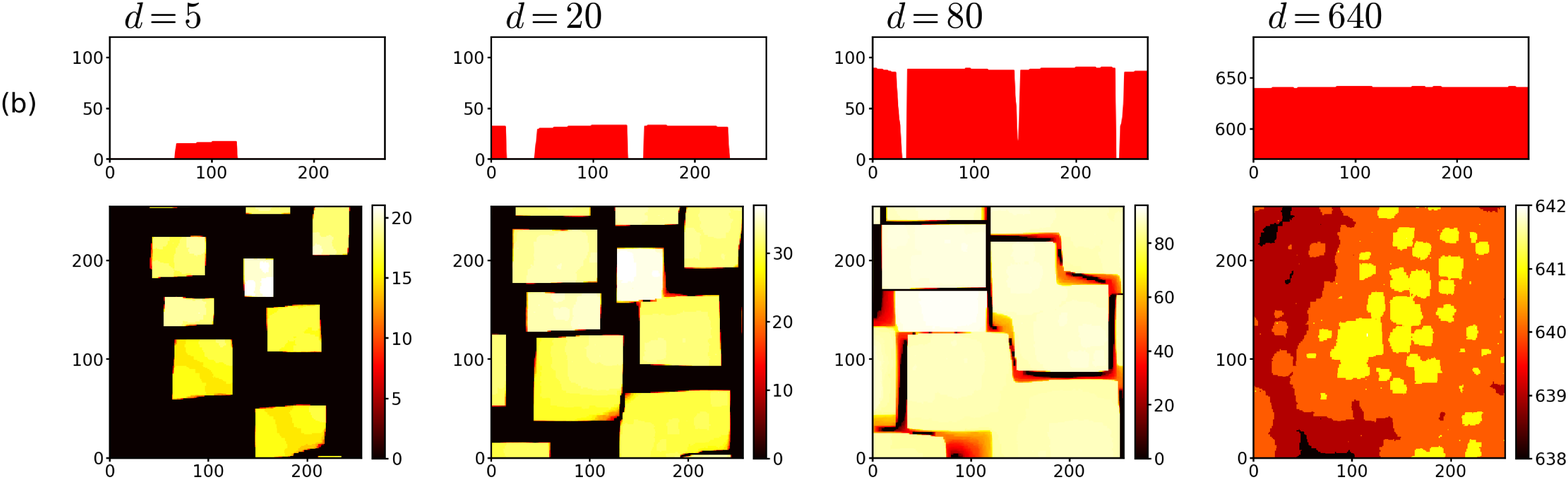}
\caption{
Top views and vertical cross sections of parts of the films grown in simulations
without ES barrier ($P=1$).
Parameter sets are (a) F and (b) M.
Cross sections are not in scale and some of them are restricted to the upper layers.
All lengths are in units of the lattice constant.
}
\label{figP1}
\end{figure}

The transition from island coalescence to continuous film growth for $E_{ES}=0$ ($P=1$) is 
illustrated in Figs. \ref{figP1}(a) and \ref{figP1}(b) for the parameter sets F and M,
respectively.
In the former, the small islands have already coalesced with $d=5$ deposited layers,
forming a large cluster with narrow gaps.
For $d=20$, a mounded topography is reminiscent of the island pattern, but for $d\geq80$
the surface has no feature of the initial islands; in the largest thickness, it has large
terraces with disordered borders.
The same features are observed in other cases where small islands are formed.
In Fig. \ref{figP1}(b), the islands are larger, taller, and more distant from each
other, so the coalescence is only observed in the panel of $d=80$.
Subsequently, the gaps between the islands are rapidly filled and there is no
evidence of their formation when $d=640$.
For this thickness, the surface has large terraces with smooth borders, as a consequence of
large $R_A$ and $\epsilon$.
Comparison with Fig. \ref{figPsmall}(b) (where the growth parameters are the same, except $P$)
shows that a smaller ES barrier delays the island coalescence.

\begin{figure}[!h]
\center
\includegraphics[height=0.21\textwidth]{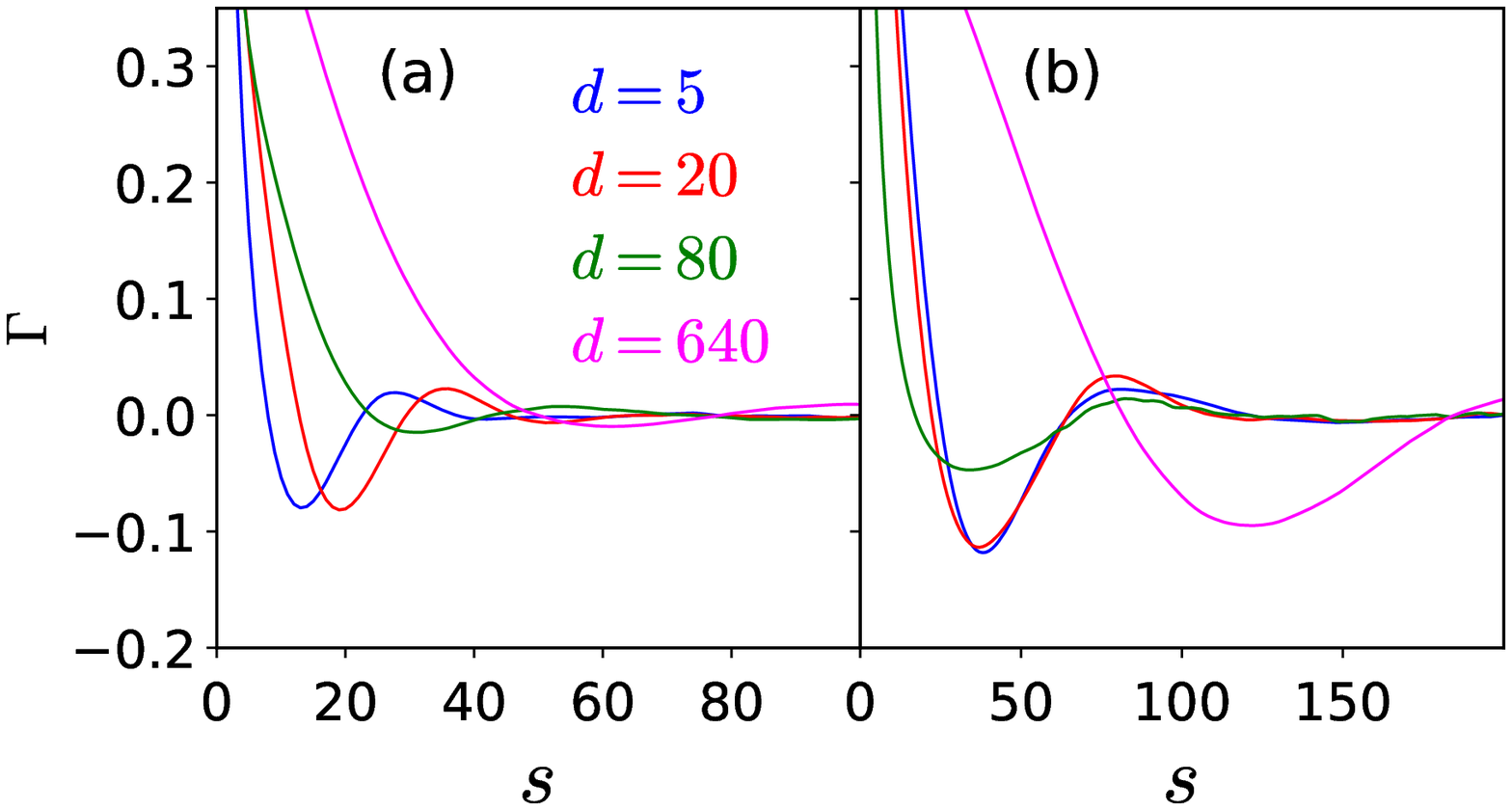}
\includegraphics[height=0.21\textwidth]{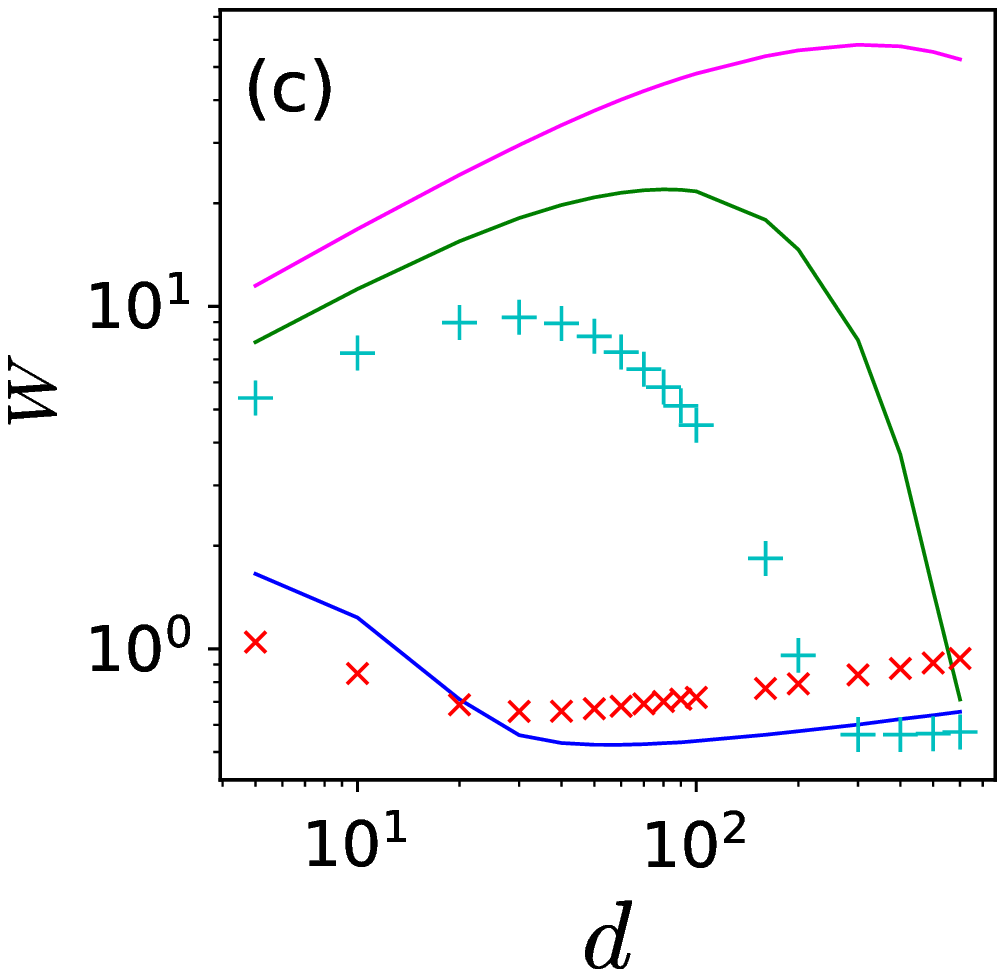}
\includegraphics[height=0.21\textwidth]{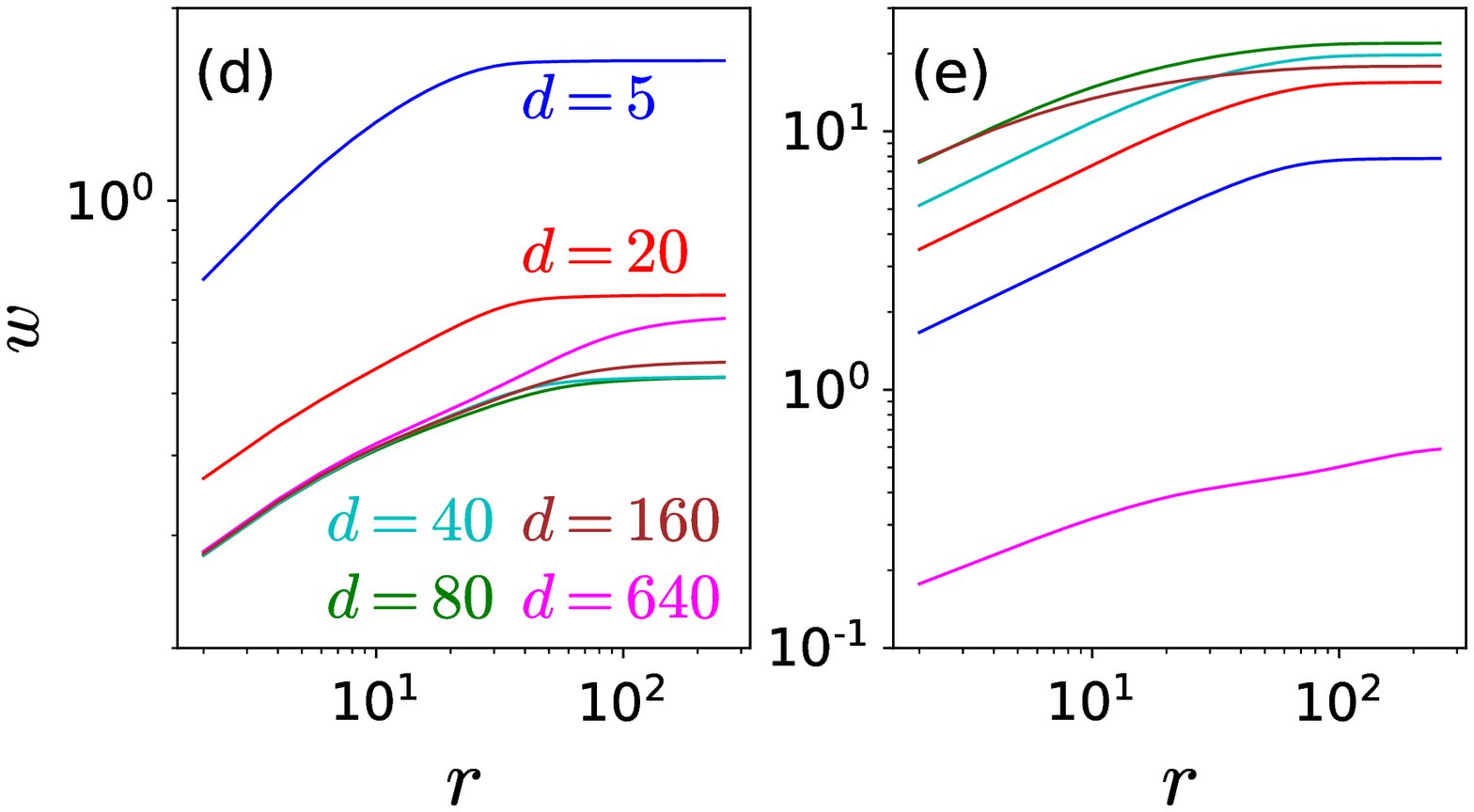}
\caption{
Auto-correlation function [(a),(b)], global roughness [(c)], and local roughness [(d)-(e)]
of films grown in simulations without ES barrier ($P=1$).
Parameter sets are: (a),(d) F; (b),(e) M;
(c) C (red crosses), F (blue line), I (cyan plus signs), M (green line), and P (magenta line).
All lengths are in units of the lattice constant.
}
\label{P1}
\end{figure}

Figures \ref{P1}(a) and \ref{P1}(b) show the auto-correlation function as a function
of the distance $s$ for the same parameters and film thicknesses of Figs. \ref{figP1}(a)
and \ref{figP1}(b), respectively.
For $d\leq20$, $\Gamma$ has a deep minimum in both cases and $s_M$ is nearly half of
the average island size.
In the smallest mobility case, a shallow minimum of $\Gamma$ is observed for $d\geq80$,
which is consistent with the absence of surface patterns.
In the largest mobility case, $\Gamma$ has no minimum at $d=80$,
but for $d=640$ it displays a minimum at $s\approx120$, which accounts for the
terraces shown in Fig. \ref{figP1}(b) (whose borders approximately follow the lattice
directions).

Figure \ref{P1}(c) shows the global roughness $W$ as a function of the thickness $d$
for five parameter sets.
In all cases, $W$ has a maximum at the thickness in which most of the islands have already 
coalesced but the gaps between them were not filled yet.
When small islands are formed, the maximum of $W$ is shown
at $d=5$ [note that Fig. \ref{P1}(c) has data only for $d\geq5$].
For the parameter set of Fig. \ref{figP1}(b), the maximum is at $d\approx80$.
The initial rapid growth of $W$ is a consequence of the rapid increase of the island heigths,
which is possible because the interlayer transport is facilitated.
The subsequent decrease is a consequence of the formation of a continuous film, in which the
facile interlayer transport rapidly suppresses the large height differences.
This type of smoothening process is similar to that observed in high temperature deposition
on rough substrates \citep{smoothening2015}.

Figs. \ref{P1}(d) and \ref{P1}(e) show $w$ as a function of  $r$
for the parameter sets of Figs. \ref{figP1}(a) and \ref{figP1}(b), respectively, and several thicknesses.
The apparent anomalous scaling is observed in Fig. \ref{P1}(e) (large islands)
in the regime of island growth, but it ceases when the islands coalesce.
During this coalescence and formation of a continuous film,
the $\log{w}\times\log{r}$ plot is displaced down as the thickness increases,
i.e. the height fluctuations decrease at all lengthscales;
this is expected in a smoothening process.

When the global roughness decreases to a value $\sim 1$ (i.e. near one lattice constant),
the memory of the initial islands is already lost and subsequent growth occurs as in initially
flat substrates.
The roughening is that of the CV model without ES barriers \citep{cv2015,carrascoPRRes2020}
and, for $R_A\gtrsim{10}^6$, the growth is almost layer by layer in experimentally reasonable film
thicknesses \citep{mozo2020}.
For instance, in Fig. \ref{P1}(d), $w<1$ from $d=80$ to $d=640$ and the $w\times t$ plot
is consistent with the normal roughening of the CV model
with large $R_A$, in which $w$ increases only in the largest windows \citep{cv2015}.
Moreover, the initial decay of $\Gamma$ becomes slower as the film grows, which
accounts for a time increasing correlation length of the fluctuations
of a self-afine surface \citep{zhao}.
However, in Fig. \ref{P1}(e), this regime begins at $d=640$ because
island coalescence occurs at longer times.

\subsection{Film Formation with Small ES Barrier}
\label{smallES}

\begin{figure}[!h]
\center
    \includegraphics[width=\textwidth]{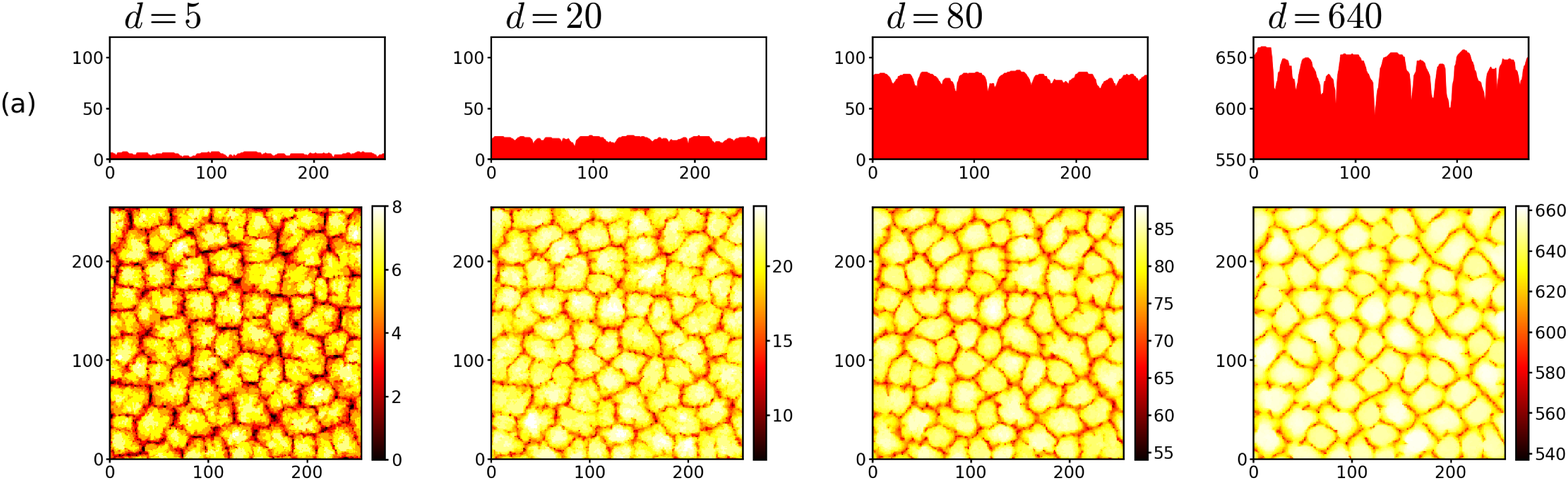} \\
    \vspace{0.5cm}
    \includegraphics[width=\textwidth]{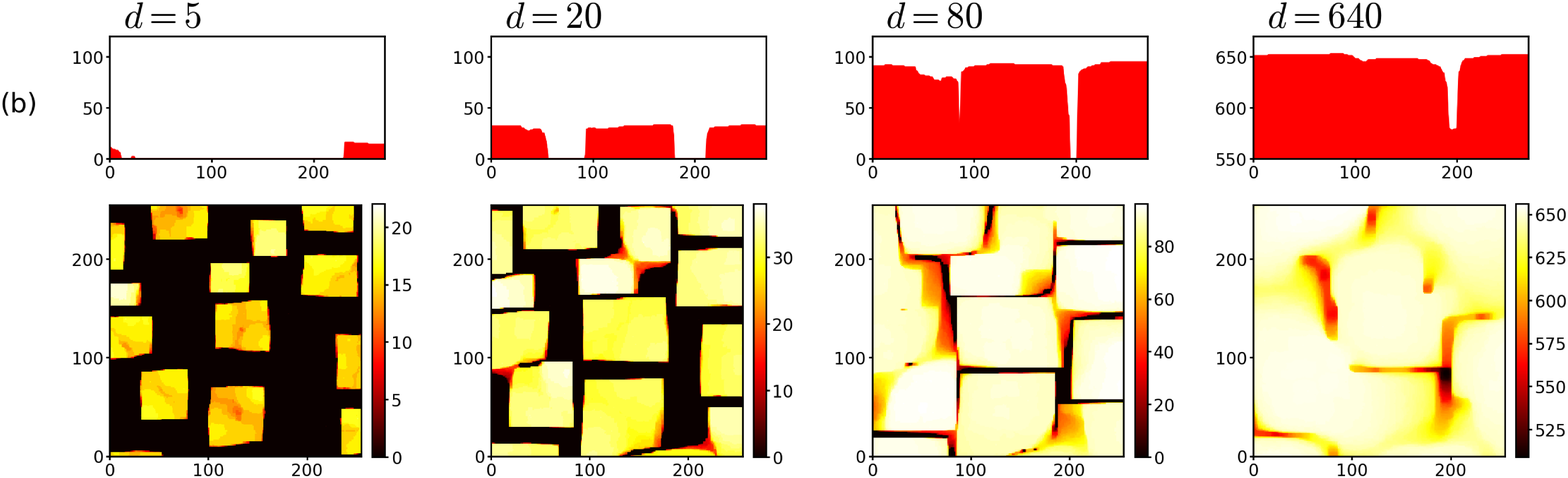}
\caption{
Top views and vertical cross sections of parts of deposits grown in simulations
with small ES barrier ($P={10}^{-1}$).
Parameter sets are (a) E and (b) L.
Cross sections are not in scale and some of them are restricted to the upper layers.
All lengths are in units of the lattice constant.
}
\label{figP01}
\end{figure}

Here we analyze the deposition with $P=0.1$, corresponding to a small ES barrier.
Figures \ref{figP01}(a) and \ref{figP01}(b) show the surfaces of the deposits for the sets
E and L, respectively, with several thicknesses.
The increase of adatom mobility also leads to
formation of larger and taller islands and a delay of their coalescence.
As in the case of large ES barrier,
the initial gaps between the islands are still visible in the thickest films,
particularly in the case of smaller adatom mobility [Fig. \ref{figP01}(a)].
However, smoother structures are observed on the top of the initial islands.

\begin{figure}[!h]
\center
\includegraphics[height=0.21\textwidth]{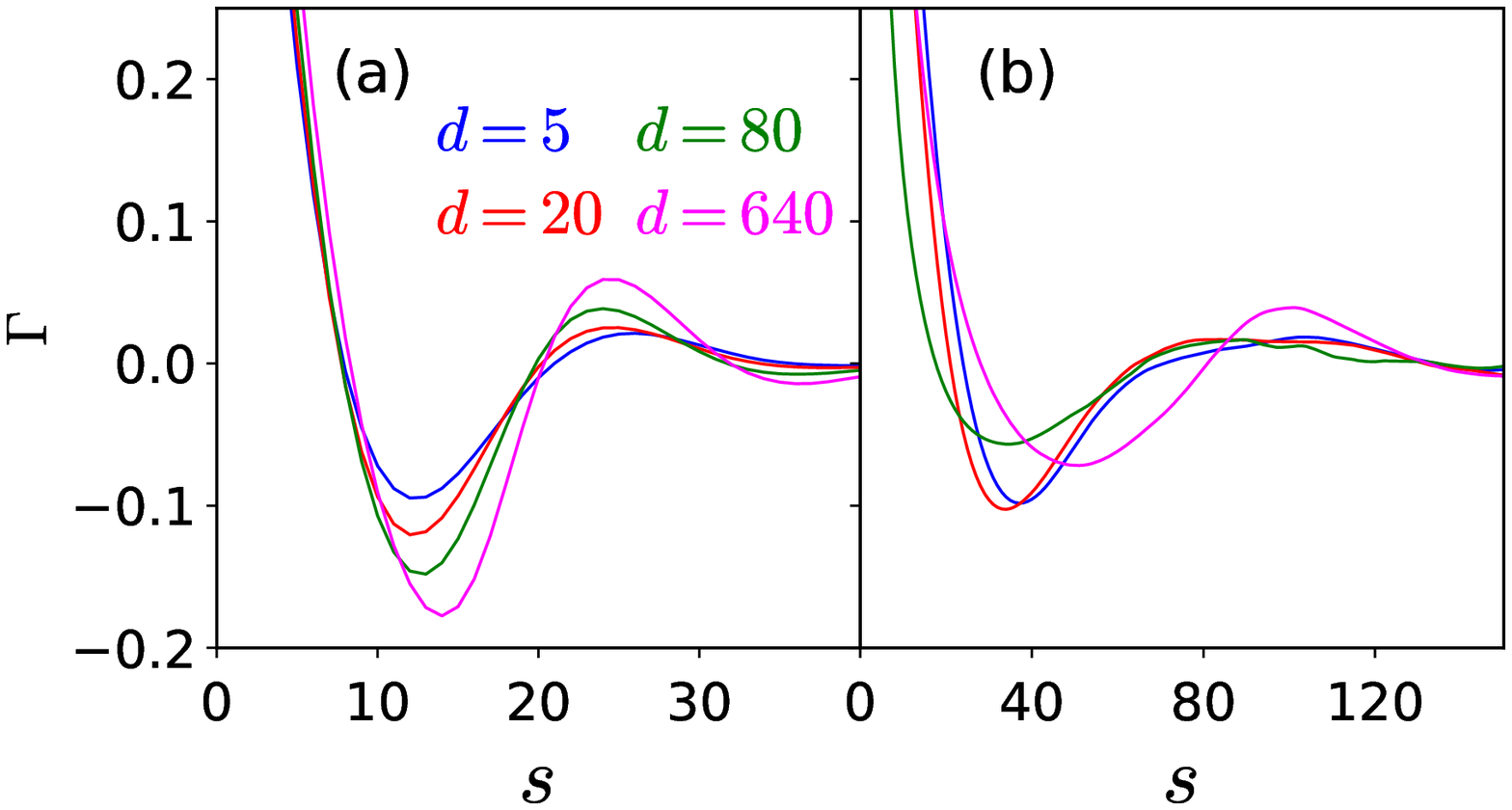}
\includegraphics[height=0.21\textwidth]{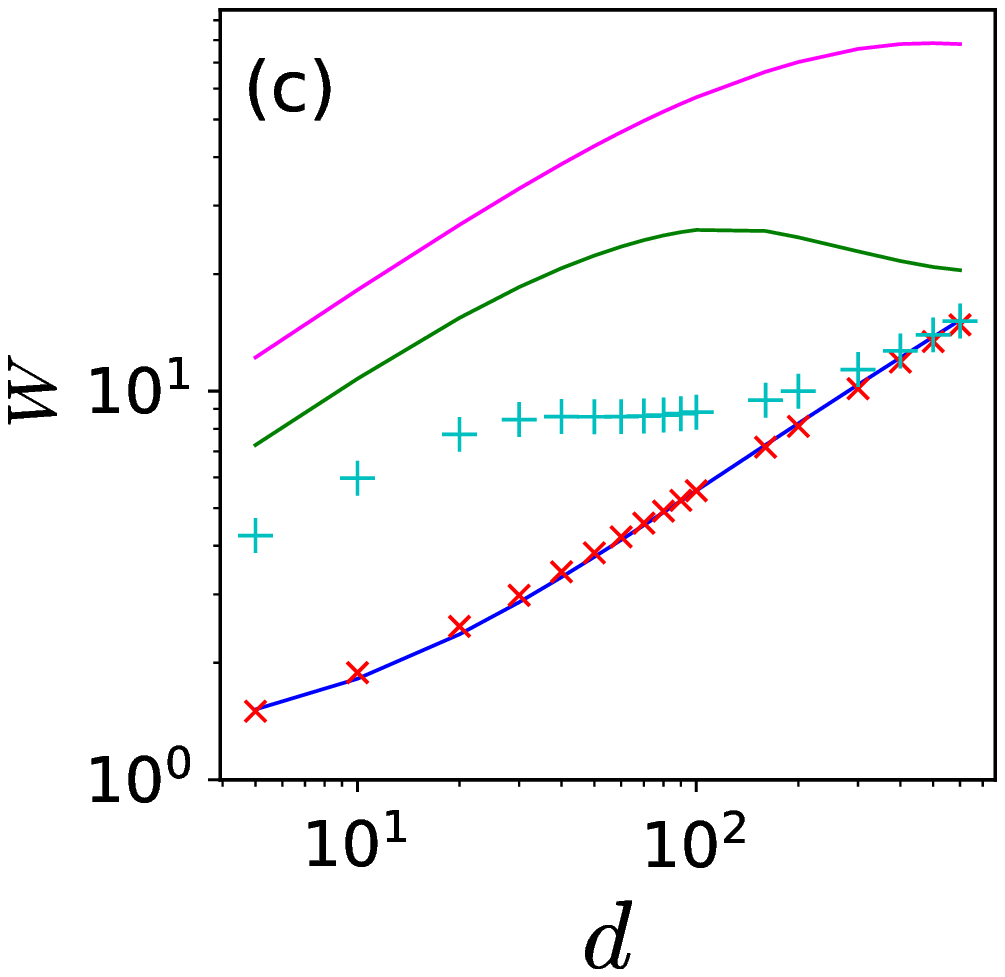}
\includegraphics[height=0.21\textwidth]{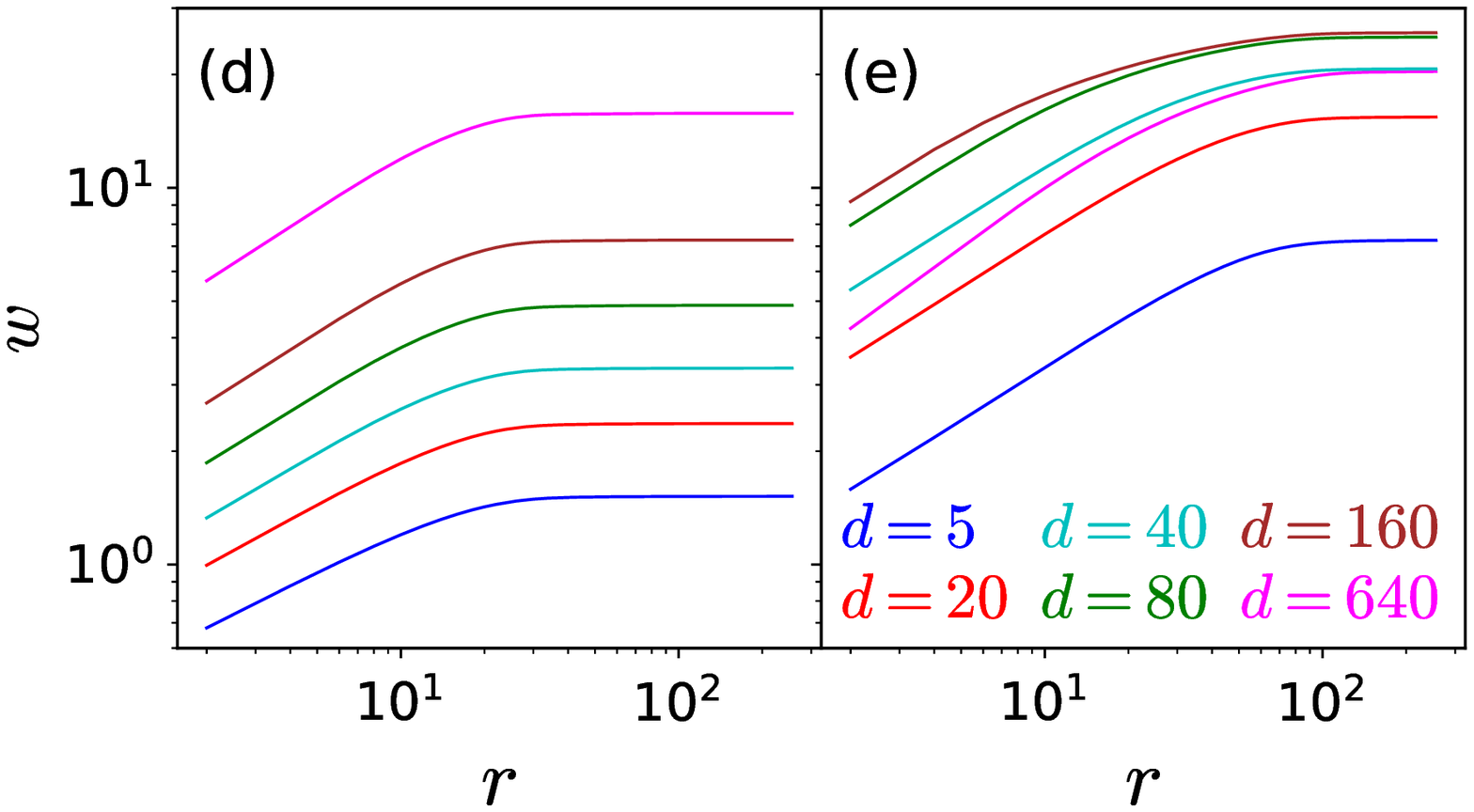}
\caption{
Auto-correlation function $\Gamma$ (a and b), global roughness (c),
and local roughness (d and e) of films grown in simulations with small ES barrier ($P=0.1$).
Parameter sets are: (a),(d) E; (b),(e) L;
(c) B (red crosses), E (blue line), H (cyan plus signs), L (green line) and O (magenta line).
All lengths are in units of the lattice constant.
}
\label{P01}
\end{figure}

Fig. \ref{P01} shows the auto-correlation function [(a), (b)],
the global roughness (c), and the local roughness [(d), (e)], for several deposition parameters
and thicknesses.

For the case of small islands, illustrated in Fig. \ref{figP01}(a),
the minimum of $\Gamma$ shown in Fig. \ref{P01}(a) has an increasing depth
but $s_M=10{\text{--}}15$ is almost constant, 
which indicates the continuous enhancement of the initial island pattern.
In Fig. \ref{P01}(d), $w$ shows an apparent anomaly, in which it increases with the
thickness for small and large $r$.
The corresponding global roughness $W$ [Fig. \ref{P01}(c)] slowly increases for small thicknesses,
but subsequently scales as $W\approx Cd^{0.5}$, with a prefactor $C$ smaller than that
obtained for large ES barrier [Fig. \ref{P001}(c)].
This means that a weaker ES barrier enhances the correlations at small scales,
but the growth is still uncorrelated at length scales larger than the island size.
If the values of $R_S$, $R_A$, or $\epsilon$ are smaller than those of Fig. \ref{figP01}(a),
we obtain results similar to those of large ES barrier, as described in Sec. \ref{largeES}.

In the case of larger islands, illustrated in Fig. \ref{figP01}(b),
there is coarsening of the structures formed on the top of the islands because the ES
barrier is small.
In the thickest films, the minimum of $\Gamma$ is shallower and $s_M$ is larger
[Fig. \ref{P01}(b)].
$W$ remains approximately constant in $d=80$ and $160$ [Fig. \ref{P01}(c)]
and decreases in the largest thicknesses.
The local roughness $w$ for small windows follows the same trend [Fig. \ref{P01}(e)].
Plateaus of $W$ are also observed for other model parameters in Fig. \ref{P01}(c);
they indicate a balance between mechanisms of roughening (typical of large $E_{ES}$)
and smoothening (typical of $E_{ES}=0$).
If $R_S$, $R_A$, and $\epsilon$ exceed the values of Fig. \ref{figP01}(b)
by factors ${10}^1{\text{--}}{10}^2$, the results are similar, although
the roughness may have a small decrease at the largest thicknesses.

The general trend observed here is that, with small adatom mobility, the ES barrier
controls the transition from island to film growth, as in the case of large ES barrier.
However, large adatom mobility may affect the transition properties.

\section{Discussion}
\label{discussion}

\subsection{A Framework to Analyze the Transition from Island to Film Growth}
\label{framework}

The island density is approximately predicted by previous results of low coverage
submonolayer growth \citep{submonorev} in terms of the diffusion coefficient of
atoms/molecules on the substrate and of their probabilities of detachment from NNs.
Larger diffusivity on the substrate generally leads to wider islands.
The ES barrier has a weak effect on the island density and, consequently, a weak effect on
the island width before coalescence.
However, since the ES barrier controls the transport from the substrate to the island tops,
a smaller $E_{ES}$ favors the formation of taller islands, which
delays the island coalescence.

If the islands are small (widths $\lesssim30$ lattice constants), the coalescence
occurs at short times and
the ES barrier is the main quantity that affects the subsequent morphology.
For large barriers, the initial island pattern remains at the surface
until long times after the continuous film formation;
the global roughness $W$ has an uncorrelated increase with the thickness ($\sim d^{1/2}$) and
the local roughness shows an apparent anomalous scaling.
Instead, with negligible ES barrier, there is a smoothening process as the islands coalesce
and a continuous film begins to grow; this is followed by a normal roughening regime.

If the islands are large,
$W$ is large at short times and reaches a plateau or a maximum as the islands coalesce.
For large ES barriers, mounds with increasing slopes grow on the top of the initial islands,
so the plateau is followed by a regime in which $W$ increases (this may take a long time
depending on the island size); the local roughness shows an anomalous behavior before
and after island coalescence.
With negligible ES barrier, $W$ has a maximum when the islands coalesce and decreases with the formation
of a continuous film (but may slowly grow at longer times).
Similar features were recently observed in Ref. \cite{empting2021PRE}.

In the regimes of island growth and coalescence, the local roughness follows the same trend of
the global roughness $W$ in small and large window sizes; this includes the cases in which
$W$ decreases with the thickness (smoothening).
The position of the first minimum of the autocorrelation function, $s_M$, may be used to
determine a characteristic lateral size of the surface patterns:
initially, it is related to the widths of the islands;
after they coalesce, if large and rounded mounds are formed, their sizes are related to $s_M$.

\subsection{The Kinetics of Low Temperature CdTe Deposition on Kapton}
\label{discussionCdTe}

The autocorrelation function in the intial stages of the CdTe film growth
[Figs. \ref{dadosCdTe}(a)] has shallow minima at $s_M\sim 100{\text{--}}150$~nm,
which correspond to island sizes $\sim 200{\text{--}}300$~nm.
Considering that the lattice constant of the CdTe crystal is $\approx 0.65$~nm and using
this value in the model, rough estimates
of the island sizes are $\sim 300{\text{--}}450$ lattice constant units.
This gives dimensionless island densities (i.e. number of islands per substrate site)
$N_{isl}\sim{10}^{-6}{\text{--}}{10}^{-5}$.
Here we consider the scaling $N_{isl}\sim {R_S}^{-3/5}$ predicted in
Ref. \protect\cite{submonorev} for $R_S$ not too large, as discussed in Sec. \ref{islandgrowth}
[i.e. a limit of Eq. (\ref{Nisl}) in which $N_{isl}$ weakly depends on $\epsilon$].
It gives $R_S\sim{10}^{8}{\text{--}}{10}^{10}$,
which is actually a condition of large mobility in our model.
Considering the deposition rate of $14.0\pm0.3$~nm/min,
the diffusion coefficient of CdTe on the Kapton surface is estimated as
${10}^{-7}{\text{--}}{10}^{-5}{\text{cm}}^2/{\text s}$ at $150\,^{\circ}\mathrm{C}$.

The variations of global and local roughness in the transition
from island to continuous CdTe film growth have many similarities with those observed
in our model with large adatom mobility (simulations with $R_S\geq{10}^8$ and
$\epsilon\geq{10}^{-2}$) but also with negligible ES barrier ($P=1$).
First, the $\log{W}\times\log{d}$ plots have maxima when the islands
are coalescing; see Figs. \ref{dadosCdTe}(b) for CdTe and \ref{P1}(c) for the model.
Second, the $\log{w}\times\log{r}$ plots are initially displaced to larger values
as the thickness $d$ increases and, after island coalescence, they are displaced to smaller values
as $d$ increases; see Figs. \ref{dadosCdTe}(c) for CdTe and \ref{P1}(e) for the model.

The kinetic roughening of much thicker CdTe films was already studied.
\citep{almeida2014,almeida2017}.
In Ref. \protect\cite{almeida2017}, the global roughness of films deposited on Kapton
fluctuate between $15$~nm and $25$~nm for thicknesses between $300$~nm and $3~\mu$m.
In films thicker than $5~\mu$m, it increases as $W\sim d^{0.24}$, which is the scaling of
the KPZ class.
The distributions of heights, local roughness, and extremal heights of those
films also agree with numerically estimated KPZ distributions.
The same KPZ scaling is also observed in thick CdTe films on Si(001) substrates \citep{almeida2014}.

KPZ scaling is a type of normal roughening, in opposition to anomalous roughening
\citep{ramasco,lopez2005}; in other words, in KPZ growth, the local roughness is time independent
in small windows (except for vanishing corrections) and increases in time in windows of the
order of the lateral correlation length or larger.
Normal roughening is not obtained with ES barriers.
For instance, Ref. \protect\cite{lealJPCM} simulated the same model studied here using $R_S=R_A$
and $E_{ES}=0.07$~eV ($P\approx 0.2$) and showed a columnar morphology, while the roughness
scaled as $W\sim d^{0.33}$.
Models of homoepitaxial growth with step edge barriers shows mound formation
\citep{etb}, in which anomalous roughening is expected.
For these reasons, the observed KPZ scaling in thick CdTe films can be explained only
if the ES barrier is very small.

Thus, the framework developed here to analyze the roughness evolution at short deposition times
gives a result consistent with the dynamic scaling applied to long deposition times.
Moreover, it becomes clear that the approach to analyze short time features is very different
from that of dynamic scaling; for instance, the displacement of $\log{w}\times\log{r}$ plots
to larger values before island coalescence cannot be interpreted as anomalous roughening,
and the subsequent smoothening is inconsistent with the usual dynamic scaling relations.

Although our model qualitatively explains the transition from island to film formation,
it cannot be used for a quantitative description of the island morphology
because it is built on a simple cubic lattice.
The model is also not suitable for a quantitative description of the
polycrystalline structures of thick CdTe films \citep{almeida2014,almeida2017}.
Despite these limitations, the compatibility with negligible ES barriers and the
estimate of the diffusion coefficient on Kapton may be important for future quantitative
modeling of vapour deposited CdTe films.

\subsection{Possible Applications to the Deposition of Other Materials}
\label{applications}

Here we show that some published experimental results on different materials
are qualitatively similar to simulation results of this work.

We begin with two recent works on thermally evaporated perovskite films.
In Ref. \protect\cite{liuMatResExp2017}, MAPbI${}_3$ films (MA=$\text{CH}_3\text{NH}_3$)
were deposited on a Si substrate whose roughness was $\lesssim1$~nm.
The first AFM image of the deposit, obtained at $130$~s, showed isolated islands with heights
$5{\text{--}}15$~nm covering part of the substrate and showed that the global roughness $W$
increased to $\approx4$~nm;
subsequently, the islands coalesced while $W$ decreased to values smaller than $1$~nm;
continued deposition to a maximal thickness of $30$~nm led to a small increase in $W$.
In Ref. \protect\cite{parveen2020}, MAPbBr${}_3$ films were deposited on SiO${}_2$ and ITO
substrates whose initial roughnesses were $\approx0.2$~nm and $1.8$~nm, respectively.
In deposition at $50^{\text o}$C, the measured roughness increased to $16{\text{--}}18$~nm when
islands partially covered the substrates (thickness $4$~nm),
decreased to $\approx10$~nm when there was significant island coalescence
(thickness $10$~nm), and then decreased to $\approx5$~nm (thickness $40$~nm).
In both cases, our model suggests that the ES barriers for diffusion
of adsorbed molecules weakly affect the island growth and the transition to a continuous film.
Moreover, the roughness evolution is similar to that of our model for relatively
low mobility on the substrate [Fig. \ref{P1}(c), $R_S\leq{10}^7$],
in which the maximal values of the roughness are
obtained at very short times, possibly before the island coalescence.

There are also experiments with different initial evolutions of the roughness and of the surface
correlations.
In the sputtering deposition of Ta films by Yang et al \citep{yangJAP2012}, the roughness slowly
varies as $W\sim d^{0.16}$ before the island coalescence, which indicates that the average island
height slowly increases, but the scaling changes to $W\sim d^{0.49}$ when a continuous film grows.
Those authors proposed a mechanistic model in which the gaps between the islands
remain after their coalescence.
In a work with a molecular beam deposition setup, Gedda et al \citep{gedda2014} grew 
cobalt phthalocyanine films on a SiO${}_2$/Si(001) substrate whose roughness was $\approx 0.3$~nm.
Their AFM images illustrate the transition from island to film growth at substrate temperature
$120^{\text o}$C.
The roughness in this regime varies as $W\sim d^{0.37}$, but the slope of the 
$\log{W}\times\log{d}$ plot increases after the continuous film is formed.
The HHCF (height-height correlation function) curves at different times are slightly split, which suggests anomalous scaling.
In both cases, our simulation results suggest the presence of large ES barriers
in the deposited materials.

Our simulations also suggest an interpretation of the results of copper electrodeposition
by Guo and Searson \citep{guoElecComm2010}.
Using a solution with sulfates, they electrodeposited disk-shaped cooper islands on RuO${}_2$ .
During the island growth, there is an increase of the global roughness and an apparent
anomaly of the local roughness, which increases in time for small and for large windows
with the same exponent.
During and after island coalescence, $w$ is approximately constant in time for all length scales,
with the global value $W$ near $600{\text{--}}700$~nm.
In the same work, the electrodeposition was performed in perchlorate solutions, which produces
hemispherical islands.
Similar evolutions of $W$ and $w$ are observed before island coalescence,
then $W$ reaches a plateau ($\approx 400$~nm) when they coalesce, and subsequently decreases to
$200{\text{--}}300$~nm.
The simulation results with large $E_S$ may explain the large island sizes
and, combined with a small ES barrier ($P={10}^{-1}$), explain the evolution
of the roughness (global and local).
The maintainance of a large roughness
up to a nominal thickness of $10~\mu$m and almost $3$~h of deposition may be interpreted
as a consequence of balanced mechanisms of smoothening and roughening, as observed in
the simulations with small (but not negligible) ES barriers [Fig. \ref{P01}(c)].

\section{Conclusion}
\label{conclusion}

Kinetic Monte Carlo simulations of a heteroepitaxial film growth model were performed
to investigate the transitions between the regimes of island growth, island coalescence,
and continuous film growth.
The global roughness, the local roughness, and the 
autocorrelation function were calculated.
From the results obtained for broad ranges of model parameters,
we developed a framework to relate the evolution of those quantities to the
microscopic properties of the growing material, independently of a dynamic scaling methods.
This framework emphasizes the interplay between the
diffusion coefficients of atoms/molecules on the substrate (which set the island width)
and the ES barrier for crossing step edges (which affects island coalescence and
film coarsening).

We also used AFM images to study those transitions in the initial stages of vapor phase deposition
of CdTe on Kapton using the hot wall technique.
When the islands coalesce, the global roughness reaches a maximal value $\sim30$~nm, which
decreases to $\sim10$~nm when a continuous film (thickness $350$~nm) is formed.
The increase and decrease of $W$ is accompanied by corresponding increase and decrease of the
local roughness at short and large windows.
The comparison with the simulation results suggests that the ES barriers are negligible
in the initial stages of CdTe film deposition.
Much thicker CdTe films show changes in crystalline grain orientation, but the observed
KPZ scaling \citep{almeida2017} is also consistent with small ES barriers, which consequently
seems to be a feature of the whole film growth.
The comparison with the simulation results also predicts a large mobility of CdTe on the Kapton
surface, with diffusion coefficient estimated as
${10}^{-7}{\text{--}}{10}^{-5}{\text{cm}}^2/{\text s}$ at $150\,^{\circ}\mathrm{C}$.

The application of our theoretical framework to analyze the deposition of other materials
was also discussed,
including perovskite, metallic, and organic films on several substrates.
This analysis distinguishes cases in which large, small, or negligible ES barriers are expected
to be present and cases of small or large mobility of the deposited species.
We expect that these results may be useful for future modeling of those materials.
We also expect that our methods can be extended to model other features in the
transition from islands to continuous films, e.g. those recently observed
in the deposition of other organic films \citep{spreitzer2019,chiodini2020}.

\section*{Acknowledgments}

The authors acknowledge support of the Brazilian agency CAPES for an interinstitutional
cooperation project (Procad 88881.068506/2014-01).
TBTT is supported by CAPES (PNPD20130933 - 31003010002P7).
RA is supported by CAPES (88887.370801/2019-00 - PrInt).
SOF is supported by CNPq (436534/2018-5 and 303153/2018-0) and FAPEMIG (APQ-00371-17).
FDAAR is supported by CNPq (305391/2018-6) and
FAPERJ (E-26/210.354/2018 and E-26/202.881/2018).

\bibliographystyle{elsarticle-num}
\bibliography{interfaces}

\end{document}